\IfFileExists{/dev/null}{\immediate\write18{sh makeLinks.sh}}{\immediate\write18{makeLinks.bat}}

\documentclass[a4paper,journal,twocolumn,10pt,final]{IEEEtran}

\makeatletter
\def\subsubsection{%
	\@startsection
	{subsubsection}                 
	{3}                             
	{\z@}                           
	{2.5ex plus 1.5ex minus 1.5ex}  
	{1ex plus .5ex minus 0ex}     
	{\normalfont\normalsize\itshape}
}
\makeatother

\usepackage[none]{hyphenat}
\usepackage[utf8]{inputenc}
\usepackage[T1]{fontenc}
\usepackage{textcomp}
\usepackage{scalefnt}
\usepackage{amsmath,amsfonts,amssymb,amsbsy,amstext}

\usepackage{mathtools}          
\usepackage{cmap}               
\usepackage[dvipsnames,svgnames,x11names]{xcolor}   
\usepackage{varwidth}

\let\originalleft\left
\let\originalright\right
\renewcommand{\left}{\mathopen{}\mathclose\bgroup\originalleft}
\renewcommand{\right}{\aftergroup\egroup\originalright}

\usepackage{graphicx}
\graphicspath{{.}{.\figs}}
\usepackage{placeins} 
\usepackage{float}
\usepackage{multirow}
\usepackage{tabularx}
\usepackage{booktabs}
\newcolumntype{C}{>{\centering\arraybackslash}X}
\newcolumntype{R}{>{\flushright\arraybackslash}X}
\newcolumntype{L}{>{\flushleft\arraybackslash}X}
\newcolumntype{P}{>{\centering\arraybackslash} p{0.5\linewidth}}

\usepackage[%
style=ieee,
backend=biber, 
url=false, 
]{biblatex}

\defbibheading{bibliography}[\refname]{\section*{#1}}

\makeatletter
\g@addto@macro{\UrlBreaks}{\UrlOrds}
\makeatother

\usepackage{acro}

\NewDocumentCommand{\acro}{m o m o}
{%
	\IfValueTF{#2}{%
		\IfValueTF{#4}{%
			\DeclareAcronym{#1}{short={#2},long={#3},#4}
		}{%
			\DeclareAcronym{#1}{short={#2},long={#3}}
		}
	}{%
		\IfValueTF{#4}{%
			\DeclareAcronym{#1}{short={#1},long={#3},#4}
		}{%
			\DeclareAcronym{#1}{short={#1},long={#3}}
		}
	}
}

\acro{2D}{two-dimensional}
\acro{3D}{three-dimensional}
\acro{4D}{four-dimensional}
\acro{6D}{six-dimensional}
\acro{1G}{first generation}
\acro{2G}{second generation}
\acro{3G}{third generation}
\acro{4G}{fourth generation}
\acro{5G}{fifth generation}
\acro{5GC}{5G core network}
\acro{5G-StoRM}{5G stochastic radio channel for dual mobility}
\acro{6G}{sixth generation}
\acro{3GPP}{3rd generation partnership project}
\acro{3GPP2}{3rd generation partnership project 2}
\acro{A2A}{air-to-air}
\acro{A2G}{air-to-ground}
\acro{AA}{antenna array}
\acro{AC}{admission control}
\acro{AD}{attack-decay}
\acro{AE}{antenna element}
\acro{AF}{amplify and forward}
\acro{ABS}{almost blank subframe}
\acro{ACF}{autocorrelation function}
\acro{ADSL}{asymmetric digital subscriber line}
\acro{AHW}{alternate hop-and-wait}
\acro{AI}{artificial intelligence}
\acro{AMC}{adaptive modulation and coding}
\acro{AP}{access point}
\acro{APA}{adaptive power allocation}
\acro{API}{application protocol interface}
\acro{ARQ}{automatic repeat request}
\acro{AT}{averaged time}
\acro{ARMA}{autoregressive moving average}
\acro{ASE}{average squared error}
\acro{ASC}{adaptive satisfaction control}
\acro{ATES}{adaptive throughput-based efficiency-satisfaction trade-off}
\acro{AWGN}{additive white gaussian noise}
\acro{BAP}{backhaul adaptation protocol}
\acro{BB}{branch and bound}
\acro{BC}{branch and cut}
\acro{BD}{block diagonalization}
\acro{BER}{bit error rate}
\acro{BF}{best fit}
\acro{BL}{bit loading}
\acro{BLER}{block error rate}
\acro{BLPC-1}{bit loading with power constraint in hop 1}
\acro{BLPC-2}{bit loading with power constraint in hop 2}
\acro{BPC}{binary power control}
\acro{BPSK}{binary phase-shift keying}
\acro{BRA}{balanced random allocation}
\acro{BS}{base station}
\acro{BSP}{\acs*{BS} placement}
\acro{BSR}{buffer status report}
\acro{CAP}{combinatorial allocation problem}
\acro{CAPEX}{capital expenditure}
\acro{CB}{contextual bandit}
\acro{CBF}{coordinated beamforming}
\acro{CBR}{constant bit rate}
\acro{CBS}{class based scheduling}
\acro{CC}{congestion control}
\acro{CCL}{common cell list}
\acro{CDF}{cumulative distribution function}
\acro{CDL}{clustered delay line}
\acro{CDMA}{code-division multiple access}
\acro{CIR}{channel impulse response}
\acro{CH}{channel hardening}
\acro{CHO}{conditional handover}
\acro{C-RAN}{cloud-based radio access network}
\acro{CL}{closed loop}
\acro{CLT}{central limit theorem}
\acro{CLPC}{closed loop power control} 
\acro{CN}{core network}       
\acro{CNR}{channel-to-noise ratio}
\acro{CPA}{cellular protection algorithm}
\acro{CPICH}{common pilot channel}
\acro{CoMP}{coordinated multi-point}
\acro{CQI}{channel quality indicator}
\acro{CRE}{cell range expansion}
\acro{CRM}{constrained rate maximization}
\acro{CRN}{cognitive radio network}
\acro{C-RNTI}{cell radio network temporary identifier}
\acro{CRRM}{centralized/common radio resource management}
\acro{CRS}{cell-specific reference signal}
\acro{CS}{coordinated scheduling}
\acro{CSI}{channel state information}
\acro{CTS}{clear to send}
\acro{CU}{centralized unit}
\acro{CUE}{cellular user equipment}
\acro{CWND}{congestion window size}
\acro{D2D}{device-to-device}
\acro{D2N}{drone-to-network}
\acro{DAA}{drone antenna array}
\acro{DAG}{directed acyclic graph}
\acro{DBS}{drone mounted base station}
\acro{DC}{dual connectivity}
\acro{DCA}{dynamic channel allocation}
\acro{DE}{differential evolution}
\acro{DF}{decode and forward}
\acro{DFT}{discrete fourier transform}
\acro{DIST}{distance}
\acro{DL}{downlink}
\acro{DMA}{double moving average}
\acro{DMRS}{demodulation reference signal}
\acro{D2DM}{\acs*{D2D} mode}
\acro{DMS}{\acs*{D2D} mode selection}
\acro{DPC}{dirty paper coding}
\acro{DRA}{dynamic resource assignment}
\acro{DSA}{dynamic spectrum access}
\acro{DSM}{delay-based satisfaction maximization}
\acro{DU}{distributed unit}
\acro{E2E}{end-to-end}
\acro{ECC}{electronic communications committee}
\acro{EDF}{earliest deadline first}
\acro{EE}{energy efficiency}
\acro{EFLC}{error feedback based load control}
\acro{EI}{efficiency indicator}
\acro{e-ICIC}{enhanced inter-cell interference coordination}
\acro{eMBB}{enhanced mobile broadband}
\acro{eNB}{evolved node B}
\acro{EXP}{exponential}
\acro{EPA}{equal power allocation}
\acro{EPC}{evolved packet core}
\acro{EPS}{evolved packet system}
\acro{E-UTRAN}{evolved universal terrestrial radio access network}
\acro{ES}{exhaustive search}
\acro{FCP}{fundamental counting principle}
\acro{FCA}{flow control algorithm}
\acro{FD}{full duplex}
\acro{FDD}{frequency division duplex}
\acro{FDM}{frequency division multiplexing}
\acro{FDMA}{frequency division multiple access}
\acro{FER}{frame erasure rate}
\acro{FIFO}{first in first out}
\acro{FF}{fast fading}
\acro{FRS}[FS]{fast-rat scheduling}
\acro{FS}{fast switching}
\acro{FSB}{fixed switched beamforming}
\acro{FST}{fixed \acs*{SNR} target}
\acro{FT}{fourier transform}
\acro{FTP}{file transfer protocol}
\acro{GA}{genetic algorithm}
\acro{GAP}{generalized assignment problem}
\acro{GAP-MQ}{generalized assignment problem with minimum quantities}
\acro{GATES}{generalized adaptive throughput-based efficiency-satisfaction trade-off}
\acro{GBR}{guaranteed bit rate}
\acro{GLR}{gain to leakage ratio}
\acro{gNB}{gNodeB}
\acro{GOS}{generated orthogonal sequence}
\acro{GP}{gaussian process}
\acro{GPL}{GNU general public license}
\acro{GPS}{global positioning system}
\acro{GRP}{grouping}
\acro{GSM}{global system for mobile communications}
\acro{GTEL}{wireless telecommunications research group}
\acro{HARQ}{hybrid automatic repeat request}
\acro{HBF}{hybrid beamforming}
\acro{HCPP}{hardcore point process}
\acro{HetNet}{heterogeneous network}
\acro{HD}{half duplex}
\acro{HH}{hughes-hartogs}
\acro{HardH}[HH]{hard handover}
\acro{HMS}{harmonic mode selection}
\acro{HO}{handover}
\acro{HOL}{head of line}
\acro{HPBW}{half power beamwidth}
\acro{HSDPA}{high speed downlink packet access}
\acro{HSR}{high-speed railway}
\acro{HSPA}{high speed packet access}
\acro{HTTP}{hypertext transfer protocol}
\acro{IAB}{integrated access and backhaul}
\acro{ICMP}{internet control message protocol} 
\acro{ICI}{intercell interference}
\acro{ICIC}{inter-cell interference coordination}
\acro{ID}{identification}
\acro{i.i.d.}{independent and identically distributed}
\acro{IETF}{internet engineering task force}
\acro{IPC}{individual power constraint}
\acro{UID}{unique identification}
\acro{IID}{independent and identically distributed}
\acro{IIR}{infinite impulse response}
\acro{ILP}{integer linear problem}
\acro{IMT}{international mobile telecommunications}
\acro{INV}{inverted norm-based grouping} 
\acro{IoT}{internet of things}
\acro{IP}{internet protocol}
\acro{IPv6}{internet protocol version 6}
\acro{IRA}{integrated resource allocation}
\acro{IRS}{intelligent reflecting surface}
\acro{ISD}{inter-site distance}
\acro{ISI}{inter symbol interference}
\acro{ISM}{industrial, scientific and medical}
\acro{ITU}{international telecommunication union}
\acro{JOAS}{joint opportunistic assignment and scheduling}
\acro{JOS}{joint opportunistic scheduling}
\acro{JP}{joint processing}
\acro{JRAPAP}{joint rb assignment and power allocation problem}
\acro{JS}{jump-stay}
\acro{JSM}{joint satisfaction maximization}
\acro{KKT}{karush-kuhn-tucker}
\acro{KPI}{key performance indicator}
\acro{LAC}{link admission control}
\acro{LA}{link adaptation}
\acro{LBS}{location based service}
\acro{LC}{load control}
\acro{LOS}{line of sight}
\acro{LP}{linear programming}
\acro{LTE}{long term evolution}
\acro{LTE-A}{\acs*{LTE}-advanced}
\acro{LTE-Advanced}{\ac{LTE-A}}
\acro{LTE-R}{\ac{LTE} railway}
\acro{LSP}{large scale parameter}
\acro{MADRDPG}{multi-agent deep recurrent deterministic policy gradient}
\acro{MeNB}{master \acs*{ENB}}
\acro{M2M}{machine-to-machine}
\acro{MAC}{medium access control}
\acro{MANET}{mobile ad hoc network}
\acro{MEDS}{method of exact doppler spread}
\acro{MC}{modular clock}
\acro{MCP}{minimal cost power}
\acro{MCS}{modulation and coding scheme}
\acro{MDB}{measured delay based}
\acro{MDI}{minimum \acs*{D2D} interference}
\acro{MDSM}{modified delay-based satisfaction maximization}
\acro{MDU}{max-delay-utility}
\acro{METIS}{mobile and wireless communications enablers for the twenty-twenty information society \acs*{5G}}
\acro{MF}{matched filter}
\acro{MFAP}{mobile femtocell access point}
\acro{MG}{maximum gain}
\acro{MH}{multi-hop}
\acro{mIAB}{mobile \ac{IAB}}
\acro{MILP}{mixed integer linear programming}
\acro{MIMO}{multiple input multiple output}
\acro{MINLP}{mixed integer nonlinear programming}
\acro{MIP}{mixed integer programming}
\acro{MISO}{multiple input single output}
\acro{MIT}{mobility interruption time}
\acro{ML}{machine learning}
\acro{MLWDF}{modified largest weighted delay first}
\acro{MME}{mobility management entity}
\acro{MMF}{max-min fairness}
\acro{mmMAGIC}{millimetre-wave based mobile radio access network for fifth generation integrated communications}
\acro{MMRP}{maximizing the minimal residual power}
\acro{MMRP-LB}{maximizing the minimal residual power with lower bound}
\acro{MMSE}{minimum mean square error}
\acro{mMTC}{massive machine-type communications}
\acro{mmWave}{millimeter wave}
\acro{MN}{master node}
\acro{MOS}{mean opinion score}
\acro{MPF}{multicarrier proportional fair}
\acro{MPRP}{maximization of the product of the residual powers}
\acro{MRA}{maximum rate allocation}
\acro{MR}{maximum rate}
\acro{MRC}{maximum ratio combining}
\acro{MRN}{mobile relay node}
\acro{MRT}{maximum ratio transmission}
\acro{MRUS}{maximum rate with user satisfaction}
\acro{MS}{mode selection}
\acro{MSE}{mean squared error}
\acro{MCG}{master cell group} 
\acro{MSI}{multi-stream interference}
\acro{MT}{mobile termination}
\acro{MTC}{machine-type communication}
\acro{IMS}{\acs*{IP} multimedia subsystem}
\acro{MTSI}{multimedia telephony services over \acs*{IMS}}
\acro{MTSM}{modified throughput-based satisfaction maximization}
\acro{MU-MIMO}{multi-user multiple input multiple output}
\acro{MU}{multi-user}
\acro{Multi-CUT}{multi-cell and multi-user and multi-tier}
\acro{NAS}{non-access stratum}
\acro{NB}{node B}
\acro{nBS}{neighbor base station}
\acro{NCL}{neighbor cell list}
\acro{NG}{next generation}
\acro{NGC}{next generation core network}
\acro{NLP}{nonlinear programming}
\acro{NLOS}{non-line of sight}
\acro{NMSE}{normalized mean square error}
\acro{NN}{neural network}
\acro{NOMA}{non-orthogonal multiple access}
\acro{NORM}{normalized projection-based grouping}
\acro{NP}{non-polynomial time}
\acro{NR}{new radio}
\acro{NRT}{non-real time}
\acro{NSA}{non-stand-alone}
\acro{NSPS}{national security and public safety services}
\acro{OBF}{opportunistic beamforming}
\acro{OFDMA}{orthogonal frequency division multiple access}
\acro{OFDM}{orthogonal frequency division multiplexing}
\acro{OFPC}{open loop with fractional path loss compensation}
\acro{O2I}{outdoor-to-indoor}
\acro{OL}{open loop}
\acro{OLPC}{open-loop power control}
\acro{OL-PC}{open-loop power control}
\acro{OM}[O\&M]{operational \& maintenance}
\acro{OMA}{orthogonal multiple access}
\acro{OPEX}{operational expenditure}
\acro{ORA}{orthogonal resource allocation}
\acro{ORB}{orthogonal random beamforming}
\acro{JO-PF}{joint opportunistic proportional fair}
\acro{OSI}{open systems interconnection}
\acro{PA}{power allocation}
\acro{PAIR}{\acs*{D2D} pair gain-based grouping}
\acro{PAPR}{peak-to-average power ratio}
\acro{P2P}{peer-to-peer}
\acro{PBCH}{physical broadcast channel}
\acro{PBS}{pico base station}        
\acro{PC}{power control}
\acro{PCI}{physical cell \acs*{ID}}
\acro{PDCP}{packet data convergence protocol}
\acro{PDF}{probability density function}
\acro{PDU}{protocol data unit}
\acro{PER}{packet error rate}
\acro{PF}{proportional fair}
\acro{PGRA}{probabilistic graph based resource allocation}
\acro{P-GW}{packet data network gateway}
\acro{PHY}{physical}
\acro{PL}{pathloss}
\acro{PLR}{packet loss ratio}
\acro{PRABE}{power and resource allocation based on quality of experience}
\acro{PRB}{physical resource block}
\acro{PROJ}{projection-based grouping}
\acro{PSD}{power spectral density}
\acro{PSDe}{positive semi-definite}
\acro{ProSe}{proximity services}
\acro{PS}{packet scheduling}
\acro{PSO}{particle swarm optimization}
\acro{PSS}{primary synchronization signal}
\acro{PTAS}{polynomial-time approximation scheme}
\acro{PTP}{point-to-point}
\acro{PZF}{projected zero-forcing}
\acro{QAM}{quadrature amplitude modulation}
\acro{QHMLWDF}{queue-HOL-MLWDF}
\acro{QoE}{quality of experience}
\acro{QoS}{quality of service}
\acro{QPSK}{quadri-phase shift keying}
\acro{QSM}{queue-based satisfaction maximization}
\acro{QuaDRiGa}{quasi deterministic radio channel generator}
\acro{RACH}{random access}
\acro{RAISES}{reallocation-based assignment for improved spectral efficiency and satisfaction}
\acro{RAN}{radio access network}
\acro{RA}{resource allocation}
\acro{RAP}{rb assignment problem}
\acro{RAT}{radio access technology}[long-plural-form={radio access technologies}]
\acro{RATE}{rate-based}
\acro{RAU}{remote antenna unit}
\acro{RB}{resource block}
\acro{RBG}{resource block group}
\acro{REF}{reference grouping}
\acro{RET}{remote electrical tilt}
\acro{RF}{radio frequency}
\acro{RL}{reinforcement learning}
\acro{RLC}{radio link control}
\acro{RLF}{radio link failure}
\acro{RM}{rate maximization}
\acro{RMa}{rural macro}
\acro{RMJ-SNR}{region of the minimum joint snr}
\acro{RMEC}{rate maximization under experience constraints}
\acro{RMSE}{root-mean-square error}
\acro{RNC}{radio network controller}
\acro{RND}{random grouping}
\acro{RNN}{recurrent neural network}
\acro{RRA}{radio resource allocation}
\acro{RRM}{radio resource management}
\acro{RSCP}{received signal code power}
\acro{RSRP}{reference signal received power}
\acro{RSRQ}{reference signal received quality}
\acro{RR}{round robin}
\acro{RRC}{radio resource control}
\acro{RSSI}{received signal strength indicator}
\acro{RT}{real time}
\acro{RTS}{request to send}
\acro{RU}{resource unit}
\acro{RUNE}{rudimentary network emulator}
\acro{RV}{random variable}
\acro{Rx}{receiver}
\acro{RZF}{regularized zero-forcing}
\acro{SA}{subcarrier assignment}
\acro{SAC}{session admission control}
\acro{sBS}{serving base station}
\acro{SC}{small cell}
\acro{SCon}[SC]{single connectivity}
\acro{SCG}{secondary cell group}
\acro{SCM}{spatial channel model}
\acro{SCS}{subcarrier spacing}
\acro{SC-FDMA}{single carrier - frequency division multiple access}
\acro{SCRV}{spatially consistent random variable}
\acro{SD}{soft dropping}
\acro{SDAP}{service	data adaptation protocol}
\acro{S-D}{source-destination}
\acro{SDPC}{soft dropping power control}
\acro{SDM}{space-division multiplexing}
\acro{SDMA}{space-division multiple access}
\acro{SE}{squared error}
\acro{SF}{shadow fading}
\acro{SMDP}{semi-markov	decision problem}
\acro{SeNB}{secondary \acs*{ENB}}
\acro{SER}{symbol error rate}
\acro{SES}{simple exponential smoothing}
\acro{S-GW}{serving gateway}
\acro{SIC}{successive interference cancellation}
\acro{SINR}{signal to interference-plus-noise ratio}
\acro{SLNR}{signal to leakage-plus-noise ratio}
\acro{SNN}{strictly non-negative}
\acro{SI}{satisfaction indicator}
\acro{SIP}{session initiation protocol}
\acro{SISO}{single input single output}
\acro{SIMO}{single input multiple output}
\acro{SIR}{signal to interference ratio}
\acro{SM}{subcarrier matching}
\acro{SMA}{simple moving average}
\acro{SMSE}{spatial-mean-square error}
\acro{SN}{secondary node}
\acro{SNR}{signal to noise ratio}
\acro{SON}{self organizing networks}
\acro{SoA}{state-of-the-art}
\acro{SoS}{sum-of-sinusoids}
\acro{SORA}{satisfaction oriented resource allocation}
\acro{SORA-NRT}{satisfaction-oriented resource allocation for non-real time services}
\acro{SORA-RT}{satisfaction-oriented resource allocation for real time services}
\acro{SP}{signal processing}
\acro{SPF}{single-carrier proportional fair}
\acro{SRA}{sequential removal algorithm}
\acro{SRB1}{signaling radio bearer~1}
\acro{SRS}{sounding reference signal}
\acro{SSB}{synchronization signal block}
\acro{SSP}{small scale parameter}
\acro{SSS}{secondary synchronization signal}
\acro{ST}{spanning tree}
\acro{STTD}{space time transmit diversity}[long-plural-form={space time transmit diversities}]
\acro{SU-MIMO}{single-user multiple input multiple output}
\acro{SU}{single-user}
\acro{tBS}{target base station}
\acro{SVD}{singular value decomposition}
\acro{TCP}{transmission control protocol}
\acro{TDD}{time division duplex}
\acro{TDM}{time division multiplexing}
\acro{TDMA}{time division multiple access}
\acro{TETRA}{terrestrial trunked radio}
\acro{TP}{transmit power}
\acro{TPC}{total power constraint}
\acro{TR}{technical report}
\acro{TTI}{transmission time interval}
\acro{TTR}{time-to-rendezvous}
\acro{TTT}{time-to-trigger}
\acro{TSM}{throughput-based satisfaction maximization}
\acro{TU}{typical urban}
\acro{TV}{television}
\acro{TVWS}{\acs*{TV} white space}
\acro{Tx}{transmitter}
\acro{UAV}{unmanned aerial vehicle}
\acro{UABSC}{user-assisted bearer split control}
\acro{UDP}{user datagram protocol}
\acro{UDN}{ultra-dense networks}
\acro{UE}{user equipment}
\acro{UBA}{\ac{UE}-\ac{BS} association}
\acro{UEPS}{urgency and efficiency-based packet scheduling}
\acro{UFC}{federal university of cear\'{a}}
\acro{ULA}{uniform linear array}
\acro{UL}{uplink}
\acro{UMa}{urban macro}
\acro{UMi}{urban micro}
\acro{UMTS}{universal mobile telecommunications system}
\acro{UPN}{user provided network}
\acro{URA}{uniform rectangular array}
\acro{URI}{uniform resource identifier}
\acro{URLLC}{ultra-reliable low-latency communications}
\acro{URM}{unconstrained rate maximization}
\acro{VBR}{variable bit rate}
\acro{VET}{variable electrical tilt}
\acro{VMR}{vehicle-mounted relay}
\acro{VR}{virtual resource}
\acro{VoIP}{voice over \acs*{IP}}
\acro{WCDMA}{wideband code division multiple access}
\acro{WF}{water-filling}
\acro{WID}{wireless infrastructure drone}
\acro{Wi-Fi}{wireless fidelity}
\acro{WiMAX}{worldwide interoperability for microwave access}
\acro{WINNER}{wireless world initiative new radio}
\acro{WLAN}{wireless local area network}
\acro{WMPF}{weighted multicarrier proportional fair}
\acro{WP}{work package}
\acro{WPF}{weighted proportional fair}
\acro{WSN}{wireless sensor network}
\acro{WWW}{world wide web}
\acro{WFQ}{weighted fair queuing}
\acro{XIXO}{(single or multiple) input (single or multiple) output}
\acro{XPR}{cross-polarization power ratio}
\acro{ZF}{zero-forcing}
\acro{ZMCSCG}{zero mean circularly symmetric complex gaussian}

\usepackage{siunitx}
\usepackage{datetime}
\usepackage{url}

\usepackage[caption=false,font=footnotesize]{subfig}

\usepackage{algorithm}
\usepackage{algorithmicx}
\usepackage{algpseudocode}

\usepackage{epstopdf}

\usepackage[hidelinks=true]{hyperref}

\usepackage[final]{changes} 
\definechangesauthor[name={Victor F. Monteiro}]{vfm}
\definechangesauthor[name={Tarcisio F. Maciel}]{tfm}
\setaddedmarkup{{\color{blue}#1}}
\setdeletedmarkup{{\color{red}\sout{#1}}}

\usepackage[referable,flushleft]{threeparttablex}
\AtBeginEnvironment{tablenotes}{\footnotesize} 

\DeclareMathAlphabet{\mathppl}{T1}{ppl}{m}{it}
\DeclareMathAlphabet{\mathphv}{T1}{phv}{m}{it}
\DeclareMathAlphabet{\mathpzc}{T1}{pzc}{m}{it}

\newcommand{\SecRef}[2][]{Section#1~\ref{#2}}
\newcommand{\FigRef}[2][]{Fig.#1~\ref{#2}}

\newcommand{\TabRef}[2][]{Table#1~\ref{#2}}

\usepackage{standalone}
\usepackage{shapepar}

\usepackage{tikz}
\usetikzlibrary{arrows}
\usetikzlibrary{positioning}
\usetikzlibrary{shapes.misc}
\usetikzlibrary{shapes.geometric}
\usetikzlibrary{shapes.symbols}
\usetikzlibrary{external}

\usetikzlibrary{shadows}
\usetikzlibrary{backgrounds}
\usetikzlibrary{shapes}
\usetikzlibrary{shapes.multipart}
\usetikzlibrary{matrix}
\usetikzlibrary{intersections}
\usetikzlibrary{fit}
\usetikzlibrary{calc}
\usetikzlibrary{chains}
\usetikzlibrary{scopes}
\usetikzlibrary{decorations.pathreplacing}
\usetikzlibrary{decorations.text}
\usetikzlibrary{arrows.meta} 
\usetikzlibrary{mindmap}
\usetikzlibrary{trees}

\usepackage{pgfplots}
\pgfplotsset{%
	width=0.95\columnwidth,
	height=0.25\textheight, 
	compat=1.14,
	compat/show suggested version=false,
	filter discard warning=false,
	tick label style={font=\footnotesize},
	label style={font=\footnotesize},
	every axis label={font=\footnotesize},
	grid=major,
	grid style={dashed,gray!30},
	cycle list shift=0,
	enlargelimits=false,
	legend style={%
		font=\footnotesize,
		legend cell align=left,
		nodes={inner xsep=2pt,inner ysep=1pt,text depth=0.15em},
	},
}
\usepackage{pgfplotstable}

\newtoggle{tikzexternalize}
 \togglefalse{tikzexternalize}

\iftoggle{tikzexternalize}{
	\immediate\write18{mkdir -p ./tikz-external-figs/}
	\usetikzlibrary{external}
	\tikzexternalize[prefix=./tikz-external-figs/,mode=list and make]
	\tikzset{external/system call={pdflatex \tikzexternalcheckshellescape -halt-on-error -interaction=batchmode -jobname "\image" "\texsource"}}
}{
}

\usepackage{titlesec}
\titlespacing*{\section}
{0pt}{0.5\baselineskip plus 0.25\baselineskip minus 0.25\baselineskip}{0.25\baselineskip plus 0.125\baselineskip minus 0.125\baselineskip}
\titlespacing*{\subsection}
{0pt}{0.5\baselineskip plus 0.25\baselineskip minus 0.25\baselineskip}{0.25\baselineskip plus 0.125\baselineskip minus 0.125\baselineskip}

\AtBeginDocument{%
\abovedisplayskip 4pt plus 2pt minus 2pt
\belowdisplayskip \abovedisplayskip
\abovedisplayshortskip 2pt plus 1pt minus 1pt
\belowdisplayshortskip \abovedisplayskip
\textfloatsep 4pt plus 2pt minus 2pt
\intextsep 4pt plus 2pt minus 2pt
\floatsep 4pt plus 2pt minus 2pt
\dbltextfloatsep 4pt plus 2pt minus 2pt
\dblfloatsep 4pt plus 2pt minus 2pt
   
}

\def\plotwidth{0.99\columnwidth}
\def\plotheight{0.4\columnwidth}

\pgfplotsset{common line style/.style={line width=1pt}}

\pgfplotsset{every axis plot post/.append style={
    every mark/.append style={scale=1.5}
}}

\pgfplotsset{common plots axis options/.style={
	every axis/.append style={
	  legend style={fill=gray!5, fill opacity=0.85, text opacity=1}
	},
	width=\plotwidth,
	height=\plotheight,
	grid=both,
	filter discard warning=false,
	tick label style={font=\footnotesize},
	label style={font=\footnotesize},
	every axis label={font=\footnotesize},
	grid=major,
	grid style={dashed,gray!30},
	cycle list shift=0,
	enlargelimits={true,abs value=1pt},
	ylabel shift = -0.5ex,
	legend style={%
		font=\scriptsize,
		legend cell align=left,
		nodes={inner xsep=2pt,inner ysep=1pt,text depth=0.15em},
	},
	}
}

\pgfplotsset{bar axis options/.style={
		common plots axis options,
		ybar=1pt,
		bar width = 3pt,
		enlarge x limits={true,abs value=5pt},
}}

\pgfplotsset{mcs axis options/.style={
		bar axis options,
		ymin = 0, ymax = 100,
		xtick=data,
		xticklabels={Total, MCS 1, MCS 2, MCS 3, MCS 4, MCS 5, MCS 6, MCS 7, MCS 8, MCS 9, MCS 10, MCS 11, MCS 12, MCS 13, MCS 14, MCS 15},
		x tick label style={
			font=\tiny,
			xshift = 1ex,
			rotate=45,
			anchor=east,
		},
}}

\pgfplotsset{common marker style/.style={
		mark repeat = 10,
		mark size = 1pt,
		mark options={solid},
}}

\pgfplotsset{only macros style/.style={
	common line style,
	common marker style,
	MediumSeaGreen,
	mark=triangle*,
}}

\pgfplotsset{macros and picos style/.style={
	common line style,	
	common marker style,
	SandyBrown,
	mark=square*,
}}

\pgfplotsset{iab no silence style/.style={
    common line style,
    common marker style,
    DodgerBlue,
    mark=*,
  }}

\pgfplotsset{iab with silence style/.style={
    common line style,
    common marker style,
    Red,
    mark=diamond*,
}}

\pgfplotsset{pedestrian style/.style={common line style, solid}}
\pgfplotsset{passenger style/.style={common line style, dashed}}
\pgfplotsset{aggregated style/.style={common line style, dotted}}

\pgfplotsset{load bar style/.style={
		DodgerBlue, fill
}}

\pgfplotsset{ack bar style/.style={
		DodgerBlue, fill
}}

\pgfplotsset{nack bar style/.style={
		Red, fill
}}

\pgfplotsset{pedestrian bar style/.style={
		DodgerBlue, fill
}}

\pgfplotsset{passenger bar style/.style={
		Red, fill
}}

\pgfplotsset{aggregated bar style/.style={
		Green, fill
}}

\def\plotsDataPath{figs/plots/data}

\begin{document}
\title{TDD frame design for interference handling in mobile IAB networks}

\author{Victor F. Monteiro$^{1}$, Fco. Rafael M. Lima$^{1}$, \deleted{\textit{Senior Member, IEEE},} Darlan C. Moreira$^{1}$, Diego A. Sousa$^{1,2}$, \\ Tarcisio F. Maciel$^{1}$, \deleted{\textit{Member, IEEE},} Behrooz Makki$^{3}$, \deleted{\textit{Senior Member, IEEE},} Ritesh Shreevastav$^{3}$ and Hans Hannu$^{3}$
\thanks{This work was supported by Ericsson Research, Sweden, and Ericsson Innovation Center, Brazil, under UFC.49 Technical Cooperation Contract Ericsson/UFC. The work of Fco. Rafael M. Lima was supported by FUNCAP (edital BPI) under Grant BP4-0172-00245.01.00/20. The work of Tarcisio F. Maciel was supported by CNPq under Grant 312471/2021-1. \\
$^{1}$ Wireless Telecommunications Research Group (GTEL), Federal University of Cear\'{a} (UFC), Fortaleza, Cear\'{a}, Brazil;  
$^{2}$ Federal Institute of Education, Science, and Technology of Cear\'{a} (IFCE), Paracuru, Brazil;  
$^{3}$ Ericsson Research, Sweden.} \vspace*{-0.8cm} 
}

\maketitle

\begin{abstract}
\Ac{IAB} is envisioned as a possible solution to address the need for network densification in situations where fiber connection is not \added{viable}\deleted{an option}. %
In Release 18, \ac{3GPP} will take another step on \deleted{improving}\added{extending} \ac{IAB} capabilities. %
With \ac{mIAB}, \ac{IAB} nodes can be deployed within \deleted{moving nodes}\added{vehicles}, e.g., buses. %
As \deleted{any}\added{in every} new technology, performance assessment and evaluation of the impact of interference is of interest for industry and academia. %
In this paper, we present contributions on those topics. %
\deleted{Firstly}\added{First}, we evaluate the performance \deleted{gains }of \ac{mIAB} \deleted{over}\added{compared to} fiber-connected \deleted{solutions}\added{deployments}. 
Moreover, we study the impact of interference on the performance of \ac{mIAB} \deleted{network}\added{networks} and propose a solution based on inserting silent slots on the \ac{TDD} frame pattern. %
According to our simulation results, \ac{mIAB} is capable of \deleted{considerably }improving performance of onboard \acp{UE} without harming too much the \ac{QoS} of surrounding \acp{UE}. %
Furthermore, we show that the \ac{TDD} frame pattern should be carefully designed to account for scenarios \added{with different levels of interference}\deleted{limited or not by interference}. %
\end{abstract}

\begin{IEEEkeywords}
	IAB, mobile IAB, wireless backhaul, 5G standardization, 6G, mobility, moving cell\added{, relay}.
\end{IEEEkeywords}

\section{Introduction}
\label{SEC:Intro}

Network densification has been an effective solution to the challenges faced by mobile communication systems such as the \deleted{increase in the}\added{increased} number of connected terminals, new multimedia services and high required data rates~\cite{Jungnickel2014}. %
However, the increased number of base stations per area also demands an improved backhaul infrastructure. %
Fiber is \added{the default choice, if available,}\deleted{an interesting technology} to provide backhaul connection from base stations to the core network\deleted{, if available}. %
Nevertheless, there are two main drawbacks when deploying fiber for backhaul connection~\cite{Madapatha2020, Madapatha2021}: %
i) depending on the region, the cost to deploy fiber for backhaul, e.g., trenching and installation, can be \deleted{very }high especially when dark fiber is not available; %
ii) in general, the deployment of fiber for backhaul takes time and can even be impossible \deleted{such as }in\added{, e.g.,} historical places where trenching is not an option.

\deleted{Whenever fiber could not be employed for backhaul, w}\added{W}ireless backhaul is a good \deleted{alternative}\added{complement to fiber}. %
In fact, wireless backhaul technology has been deployed in mobile networks for a long time mainly using proprietary solutions and dedicated spectrum~\cite{Edstam2018}. %
In Release 10, \ac{3GPP} standardized wireless backhaul for \ac{LTE}, the so called \ac{LTE} relaying~\cite{3gpp.overview.rel.10}. %
However, the commercial interest was not as large as expected\added{, due to, e.g., the small available spectrum resources and lack of use-case of interest at that time}. %
Motivated by the large bandwidths available in \ac{mmWave} spectrum employed in \ac{5G} networks and advanced \ac{MIMO} techniques, \ac{3GPP} began the first studies on wireless backhaul for \ac{NR} in 2017 \deleted{with}\added{under} the name of \ac{IAB}~\cite{3gpp.38.874}. %

The first set of specifications of \ac{IAB} was frozen on July 2020 in Release 16~\cite{3gpp.38.300c}. %
Optimization of some functionalities and increased efficiency are the focus of Release 17 for \ac{IAB} that should have its stage 3 protocol frozen on \deleted{March}\added{early} 2022. %
In December 2021, \ac{3GPP} presented the package approval for Release 18 setting the cornerstone for \ac{5G} advanced\deleted{ where \ac{IAB} will take one step further}. %
Under the name of \ac{mIAB}, a work item will specify protocols and \added{procedures}\deleted{architecture} needed to provide wireless backhaul under \ac{IAB} to moving nodes~\cite{3gpp.RP-213469}. %

In this paper, we \deleted{firstly }show that \ac{mIAB} networks is a \deleted{promising}\added{possible} solution to improve \ac{QoS} for \deleted{onboard \acp{UE}}\added{\acp{UE} onboard of busses}\added{,} \deleted{when }compared to classical solutions such as the deployment of fiber-based macro/pico cells. %
Furthermore, we study an \ac{mIAB} network over different interference conditions and transmission configurations. %
More specifically, we assume two distinct system layouts: one limited and another non-limited by interference. %
Furthermore, we assume different \ac{TDD} frame patterns where backhaul and access transmissions can or cannot take place simultaneously. %
Our objective is to evaluate {\it when and over which  \ac{mIAB} transmission configurations} the interference level becomes unbearable. %

\deleted{The remainder of this paper is organized as follows: Section~\ref{SEC:Background} briefly describes \ac{IAB} architecture and characterizes the main interference sources in this network; Section~\ref{SEC:Proposal} presents the state-of-the-art solutions for interference avoidance/management in \ac{mIAB} networks; Section~\ref{SEC:Perf_Eval} provides a performance evaluation for \ac{mIAB} over different scenarios with distinct interference levels and \ac{TDD} frames, and, finally, Section~\ref{SEC:Conclusion} presents the main conclusions and perspectives of this work.} %

\section{Background}
\label{SEC:Background}

\subsection{\Acl{IAB}}
\label{SUBSEC:Background_IAB}

\begin{figure}[t]
	\centering
	\includegraphics[width=0.95\columnwidth]{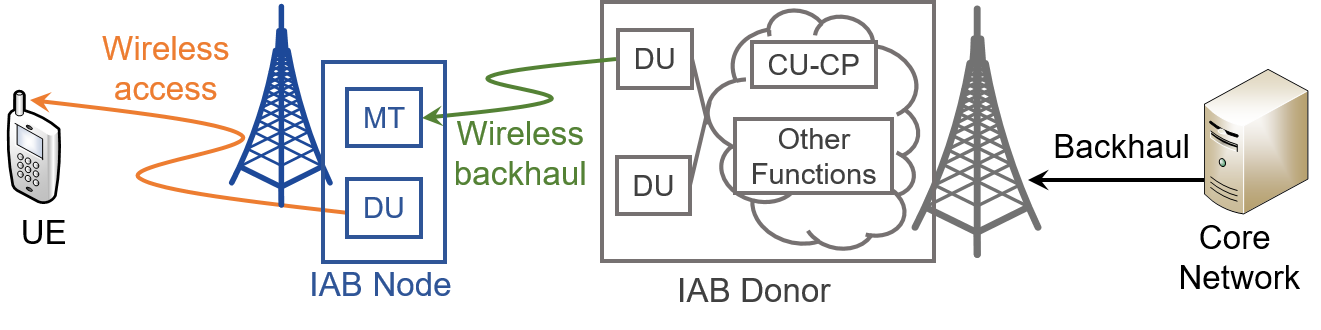}
	\caption{\added{Example of }\acs{IAB} \acs{NR} architecture.}\label{FIG:Architecture}
\end{figure}

As illustrated in \FigRef{FIG:Architecture}, an \ac{IAB} network has two main components: \ac{IAB} donor and \ac{IAB} node. The \ac{IAB} donor \deleted{is a wired-backhauled \ac{gNB} that }provides a wireless backhaul to \ac{IAB} nodes allowing these to provide access to their served \acp{UE}. %

Concerning the \ac{IAB} donor, it is split in \deleted{\acp{CU}}\added{\ac{CU}} and \acp{DU}. %
This split is transparent to the served \deleted{nodes}\added{\acp{UE}} and these units can be either collocated or non-collocated. %
In the \ac{IAB} donor, the \ac{DU} terminates lower protocol layers, e.g., \ac{PHY}, \ac{MAC}, and \ac{RLC}, while the \ac{CU} terminates upper protocol layers, e.g., \ac{PDCP} and \ac{RRC}. %
\added{The motivation for the Rel-16 chosen architecture was centralization (not having a \ac{CU} in every IAB-node), equivalent to lean \ac{IAB} nodes.} %
\deleted{The main objective of this split is to allow time-critical functionalities, e.g., scheduling and retransmission, to be performed in \acp{DU} closer to the served nodes, while other functionalities can be performed in \deleted{\acp{CU}}\added{\ac{CU}} with better processing capacity. }%

Regarding the \ac{IAB} nodes, they support \ac{gNB} \ac{DU} and \ac{MT} \deleted{functionalities}\added{functions}, as illustrated in \FigRef{FIG:Architecture}. 
On the one hand, the \added{\ac{IAB}-}\ac{MT} \deleted{part of an \ac{IAB} node }terminates the radio interface layers of its backhaul towards an upstream \ac{IAB} donor\added{ or parent \ac{IAB}}. %
On the other hand, the \added{\ac{IAB}-}\ac{DU}\deleted{ part} terminates the \ac{NR} interface to \acp{UE} \added{ and/or child \ac{IAB} nodes}. %
For compatibility purposes with legacy networks, the \added{\ac{IAB}-}\ac{MT} \deleted{part of an \ac{IAB} node} acts \deleted{as a regular}\added{not different from a} \ac{UE} from the point-of-view of its \added{parent }\ac{IAB}\deleted{ donor, i.e., its serving \ac{gNB}}. %
From a \ac{UE} point of view, the \added{\ac{IAB}-}\ac{DU} \deleted{part of an \ac{IAB} node }looks like as the \ac{DU} of a regular \ac{gNB}. %

As already mentioned, in \ac{3GPP} Release~$18$, \deleted{it is expected that }different \ac{RAN} \added{working }groups will work towards enhancing functionalities of \ac{IAB} focusing on \ac{mIAB} to provide \ac{5G} \added{services}\deleted{coverage enhancement} to onboard \deleted{and surrounding }\acp{UE}. %
\deleted{The initial use cases are expected to be based on~\cite{3gpp.22.839}. }%
One of the main use-cases of \ac{mIAB} cells is to serve \acp{UE} onboard of\added{, e.g., busses}\deleted{ vehicles}. %
Some advantages of \ac{mIAB} are to improve connectivity to the network by reducing/eliminating the vehicle penetration loss (especially at high frequency) and to avoid signalling storms from simultaneous \ac{HO} messages. %
\deleted{Other relevant use cases involve nomadic \ac{IAB} network nodes mounted on vehicles to provide extended coverage to surrounding \acp{UE}. %
Nomadic \ac{IAB} nodes typically operate while being stationary in specific places, e.g., close to events, disaster areas, etc. }%
%

\subsection{Interference in \ac{mIAB} Scenario}
\label{SUBSEC:Background_inter_mng}

A challenge that appears in scenarios with \ac{mIAB} is the interference management. %
Two major types of interference in these scenarios are the self-interference and the dynamic interference between \added{mobile and legacy deployed stationary}\deleted{moving cells and crossed fixed} cells that may occur when \deleted{these cells}\added{they} \added{share}\deleted{use the same} frequency spectrum. %
These two types of interference are explained in the following. %

\begin{figure}[t]
	\centering
	\includegraphics[width=0.95\columnwidth]{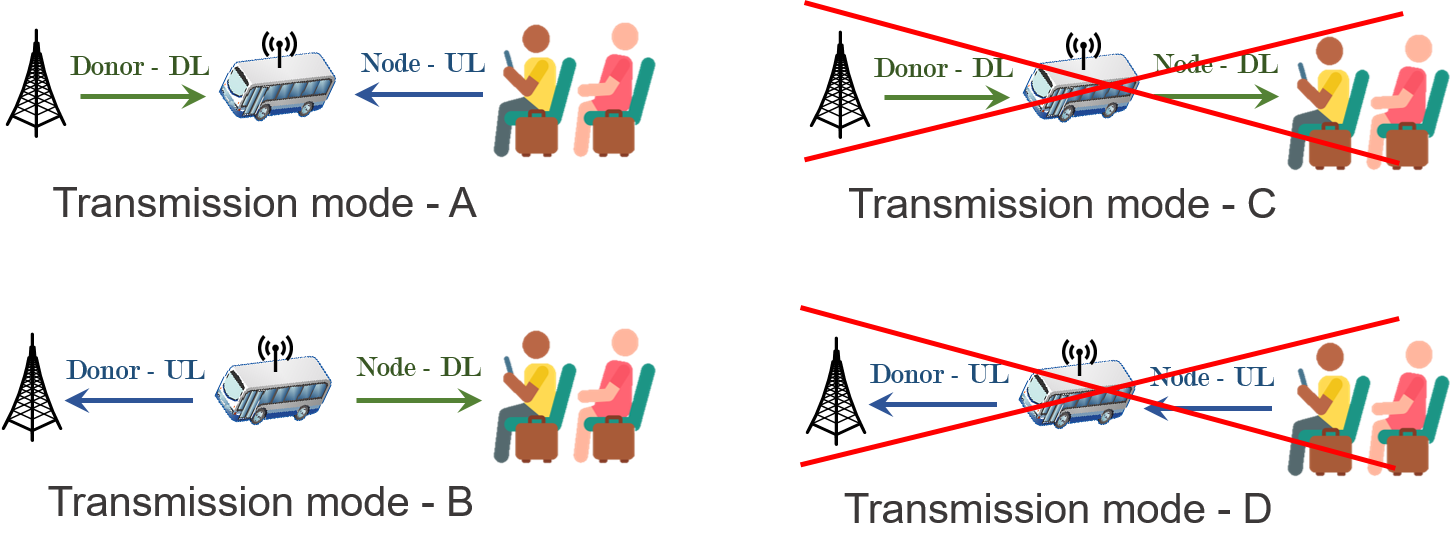}
	\caption{List of possible simultaneous operation modes.}
	\label{FIG:Transmission-modes}
\end{figure}

Figure~\ref{FIG:Transmission-modes} presents $4$~possible simultaneous transmission modes in which an \ac{mIAB} can operate. %
In modes A and B, the \ac{DU} and \ac{MT} parts of an \ac{mIAB} node perform the same actions (either receive or transmit data), while in modes C and D they perform opposite actions. %
Modes C and D are often referred to as \ac{IAB} \ac{FD} modes, where the \ac{MT}-part of the \ac{mIAB} node receives data in the backhaul while its \ac{DU}-part transmits data to access \ac{UE}, or vice versa. %
In \ac{IAB} \ac{FD} modes, the part which is transmitting may cause strong interference to the part that is receiving, which is the so called self-interference. %

Self-interference can only be mitigated in very specific \added{implementations and }scenarios, e.g., where both \ac{DU} and \ac{MT} parts of the \ac{mIAB} node are very isolated from each other or complex signal processing strategies in analog and digital domains are employed. %
Thus, in order to avoid the problem of self-interference, the network usually allows an \ac{mIAB} node to only operate in transmission modes A and B, resulting in the \ac{TDD} scheme showed in \TabRef{TABLE:TDD-self-interf} (for an \ac{IAB} network operating with a maximum of two hops, i.e., where an \ac{mIAB} node cannot be served by another \ac{mIAB} node). %

%
\begin{table}[!t] 
	\centering
	\caption{\acs{TDD} scheme avoiding self-interference.}
	\label{TABLE:TDD-self-interf}
  {\setlength{\tabcolsep}{3pt}
  \begin{tabularx}{\columnwidth}{l|cc|CC|C}
		\toprule
		Slot & 1  & 2 & DL usage & UL usage & Total usage \\ 
		\midrule
		\ac{IAB} donor\added{ access} & DL & UL & 50\% & 50\% & 100\% \\
		\added{\ac{mIAB} node backhaul} & DL & UL & 50\% & 50\% & 100\% \\
		\ac{mIAB} node\added{ access} & UL & DL & 50\% & 50\% & 100\% \\
		\bottomrule
	\end{tabularx}
  }
\end{table}



\begin{figure}[t]
	\centering
	\includegraphics[width=0.8\columnwidth]{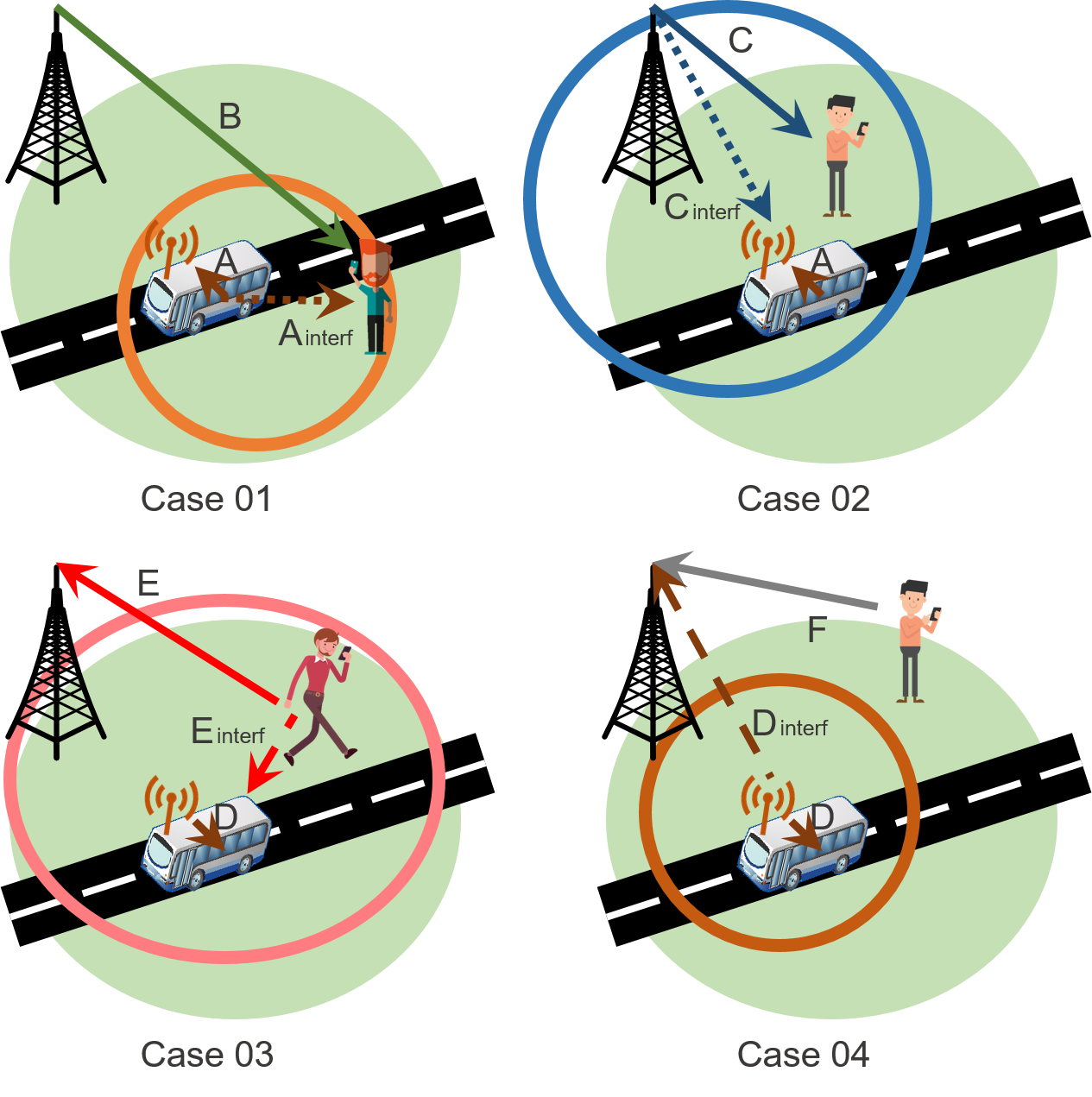}
	\caption{Examples of interference in \ac{mIAB} scenario.}
	\label{FIG:Interference-cases}
\end{figure}

The \ac{TDD} scheme in \TabRef{TABLE:TDD-self-interf} may cause dynamic interference between \ac{mIAB} cells and crossed fixed cells. %
Figure~\ref{FIG:Interference-cases} illustrates the two cases (Cases 01 and 02) of interference that occur when the \ac{IAB} donor is in \ac{DL} and the \ac{mIAB} node \deleted{is }in \ac{UL}, and the two cases (Cases 03 and 04) when the \ac{IAB} donor is in \ac{UL} and the \ac{mIAB} node \deleted{is }in \ac{DL}. %


More specifically, in Case 01, a pedestrian receiving data in the \ac{DL} (link B) from an \ac{IAB} donor may suffer interference (link $A_{\text{interf}}$) from an in-vehicle passenger transmitting in the \ac{UL} (link A) to an \ac{mIAB} node deployed inside a bus. %
In Case 02, the \ac{DU} part of an \ac{mIAB} node receiving data in the \ac{UL} (link A) from a passenger may suffer interference (link $C_{\text{interf}}$) from the \ac{IAB} donor when it transmits in the \ac{DL} (link C) to a pedestrian. %
In Case 03, a passenger receiving data from an \ac{mIAB} node in the \ac{DL} (link D) may suffer interference (link $E_{\text{interf}}$) from a pedestrian transmitting in the \ac{UL} to its serving \ac{IAB} donor (link E). %
In Case 04, an \ac{IAB} donor receiving data in the \ac{UL} (link F) from a pedestrian may suffer interference (link $D_{\text{interf}}$) from the \ac{DU} part of an \ac{mIAB} node transmitting in the \ac{DL} (link D). %

\section{Problems with Existing Solutions and Perspectives}
\label{SEC:Proposal}

\begin{table*}[!t] 
	\centering
	\caption{\acs{TDD} scheme with silent slots.}
	\label{TABLE:TDD-silent-slots}
	\begin{tabularx}{\textwidth}{l|CCCCCCCCCC|cc|c}
		\toprule
		Slot      & 1             & 2             & 3             & 4             & 5             & 6             & 7             & 8             & 9             & 10            & DL usage & UL usage & Total usage \\  
		\midrule
		IAB donor\added{ access} & DL            & UL            & - & DL            & - & UL            & DL            & - & UL            & DL            & 40\%     & 30\%     & 70\%  \\
		\added{IAB node }backhaul  & DL            & - & UL            & DL            & UL            & - & - & UL            & - & DL            & 30\%     & 30\%     & 60\%  \\
		IAB node\added{ access}  & - & UL            & DL            & - & DL            & UL            & DL            & DL            & UL            & - & 40\%     & 30\%     & 70\% \\
		\bottomrule
	\end{tabularx}
\end{table*}

The dynamic interference between moving cells and \deleted{crossed }fixed cells can limit the \added{network }performance\deleted{ of \ac{mIAB} networks}. %
Some solutions can be envisaged to tackle this problem. %

A first option is to assume that the \ac{mIAB} network operates with two distinct frequency bands: one to serve onboard \acp{UE} and another to serve surrounding \acp{UE}. %
For example, short range \ac{mmWave} spectrum could be assigned to onboard \acp{UE} while sub-6 GHz could be used to serve surrounding \acp{UE}. %
The main disadvantage of this solution is that the load in \ac{mIAB} can be very dynamic and, in some cases, the bandwidth available in \ac{mmWave} spectrum may remain unused, e.g., when the load offered by onboard \acp{UE} is low. %

An alternative to this inflexible frequency partitioning is to employ dynamic time-frequency resource allocation where the whole bandwidth would be available to the time-frequency schedulers. %
In this case, \deleted{donor }\ac{IAB}\added{ donor} and \ac{mIAB} nodes could coordinate their transmissions in order to dynamically control the interference level in the system. %
The schedulers can take their decisions based, for example, on interference information and on the system load. %
Although we expect that this solution can substantially improve the system performance, this would be obtained at the cost of increased complexity and signalling load. %
More specifically, \added{up-to-date }system information such as interference and load measurements should be available at nodes where decision would be taken. %
Moreover, those measurements should be frequently collected, since they can vary a lot in such a dynamic network. %

Thus, in this paper we evaluate the potential of a different alternative to deal with interference in \ac{mIAB} network. %
In fact, in order to control the interference between the links as shown in \FigRef{FIG:Interference-cases}, a \ac{TDD} frame different from the one shown in Table \ref{TABLE:TDD-self-interf} can be designed. %
More specifically, silent slots can be inserted in the \ac{TDD} frame structure as, e.g., shown in Table~\ref{TABLE:TDD-silent-slots}. %
Note that in this \ac{TDD} frame structure, in some slots either the link involving \ac{IAB} donor access, \ac{mIAB} node access or backhaul is \deleted{in silence}\added{disabled}\added{, i.e., no transmission/reception of data is allowed. %
These slots are	called here in as \textit{slots of silence}}. %
Therefore, some of the interference cases shown in \FigRef{FIG:Interference-cases} can be avoided. %
For example, in Slot 1, as \ac{mIAB} node access is \added{disabled}\deleted{in silence}, the interference Case 01 in \FigRef{FIG:Interference-cases} is not an issue anymore, as the pedestrian \ac{UE} would not receive interference from \ac{UL} transmissions from \acp{UE} connected to the \ac{mIAB}. %
In Slot 3, for example, as \ac{IAB} donor access is \added{disabled}\deleted{in silence}, interference Case 03 does not hold since onboard \acp{UE} connected to the \ac{mIAB} would not experience interference from pedestrian \acp{UE} (connected in \ac{UL} to the \ac{IAB} donor). %


There is a clear trade-off when employing silent slots for interference handling. %
On the one hand, a lower interference is experienced in the whole system as previously explained. %
This leads to an improved signal level which can be translated in the use of higher-order \acp{MCS}. %
On the other hand, as some of the links are blocked depending on the slot, the overall \ac{TDD} frame efficiency decreases. %
For example, switching from the \ac{TDD} frame shown in \TabRef{TABLE:TDD-self-interf} to the one in \TabRef{TABLE:TDD-silent-slots}, the active period of the \ac{IAB} donor access drops from $100\%$ to $70\%$. 
In this case, there \deleted{is}\added{are} less opportunities to transmit which can lead to reduced throughput. %
Thus, in \added{\SecRef{SEC:Perf_Eval}}\deleted{the next section} we will evaluate the impact of the use of \ac{TDD} frame patterns with and without \deleted{silent }slots\added{ of silence} in \added{scenarios with different levels of interference}\deleted{two distinct scenarios: scenarios limited and non-limited by interference}. %

\section{Performance Evaluation}
\label{SEC:Perf_Eval}

This section presents a performance comparison between a scenario with \ac{mIAB} and two benchmark scenarios, i.e., a scenario with only macro \acp{gNB}, called here as \textit{only macros} scenario and other with macro and pico \acp{gNB} fiber-connected to the \ac{CN}, called here as \textit{macros-picos} scenario. %
The details concerning the considered simulation modelling are presented in \SecRef{SEC:Sim_Assump} and the results are discussed in \SecRef{SUBSEC:Sim_Results}. %

\subsection{Simulation Assumptions}
\label{SEC:Sim_Assump}
\begin{figure}[!t]
	\centering
	
	\subfloat{%
		\includegraphics[width=0.75\columnwidth]{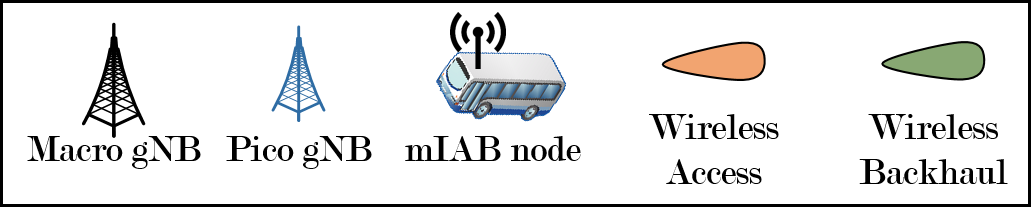}
	}
	
	\setcounter{subfigure}{0}
	
	\subfloat[Not limited by interference.]{
		\includegraphics[width=0.46\columnwidth]{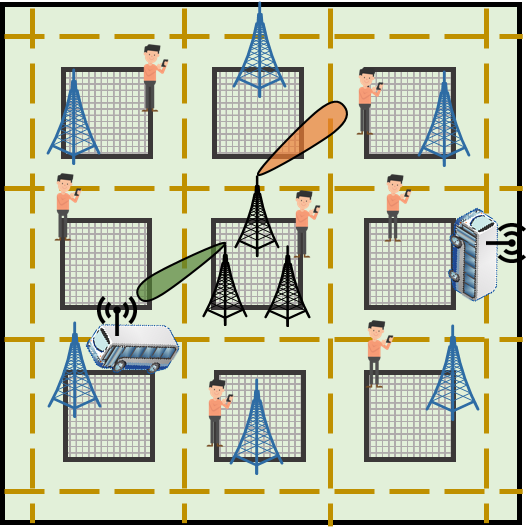}
		\label{FIG:Layout-not-limited-by-interf}
	}	
	\subfloat[Limited by interference.]{
		\includegraphics[width=0.53\columnwidth]{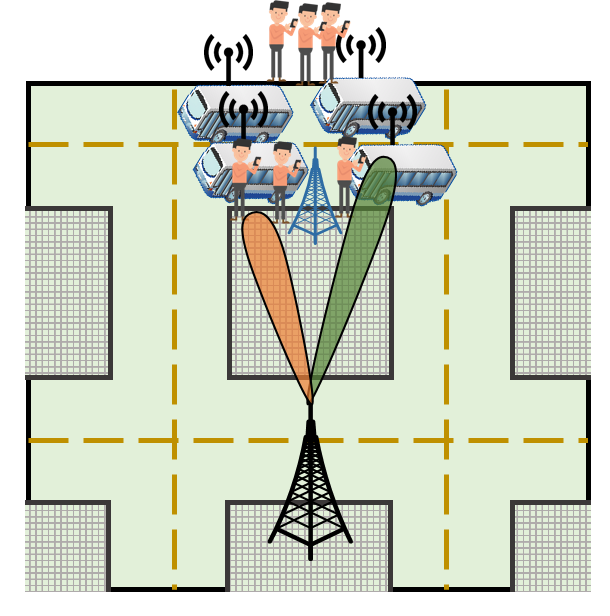}
		\label{FIG:Layout-limited-by-interf}
	}	
	\caption{Scenarios of interest.}
	\label{FIG:Simulation-scenarios}	
\end{figure}

\begin{table*}[!t] 
	\centering
	\small
	\caption{\acs{TDD} scheme adopted in only macros and macros-picos scenarios.}
	\label{TABLE:macro-pico-TDD}
	\resizebox{\textwidth}{!}{
		\begin{tabular}{l|llllllllll|ll|l}
			\toprule
			Slot                & 1  & 2      & 3  & 4  & 5  & 6  & 7      & 8  & 9  & 10 & DL usage & UL usage & Total usage \\ 
			\midrule
			Macro gNBs & DL & S (DL) & UL & UL & UL & DL & S (DL) & UL & UL & DL & 50\%     & 50\%     & 100\% \\
			\bottomrule
		\end{tabular}
	}
\end{table*}

\begin{table*}[!t]
	\centering
	\caption{Entities characteristics.}
	\label{TABLE:Entities-characteristics}
	\begin{tabularx}{\textwidth}{lXXXXX}
		\toprule
		\textbf{Parameter} & \textbf{\ac{IAB} Donor} & \textbf{\ac{mIAB} node - \ac{DU}} & \textbf{\ac{mIAB} node - \ac{MT}} & \textbf{Pedestrian} & \textbf{Passenger} \\
		\midrule
		Height & \SI{25}{\meter} & \SI{2.5}{\meter} & \SI{3.5}{\meter} & \SI{1.5}{\meter} & \SI{1.8}{\meter} \\
		Transmit power & \SI{35}{\decibel m} & \SI{24}{\decibel m} & \SI{24}{\decibel m} & \SI{24}{\decibel m} & \SI{24}{\decibel m} \\
		Antenna tilt & $12^{\circ}$ & $4^{\circ}$ & $0^{\circ}$ & $0^{\circ}$ & $0^{\circ}$ \\
		Antenna array & URA $8\times 8$ & URA $8\times 8$ & ULA $64$ & Single antenna & Single antenna \\
		Antenna element pattern & \ac{3GPP} 3D~\cite{3gpp.38.901} & \ac{3GPP} 3D~\cite{3gpp.38.901} & Omni & Omni & Omni \\
		Max. antenna element gain & \SI{8}{\decibel i} & \SI{8}{\decibel i} & \SI{0}{\decibel i} & \SI{0}{\decibel i} & \SI{0}{\decibel i} \\
		Speed (not limited by interference) & \SI{0}{km/h} & \SI{40}{km/h}  & \SI{40}{km/h} & \SI{3}{km/h} & \SI{40}{km/h} \\
		Speed (limited by interference) & \SI{0}{km/h} & \SI{20}{km/h}  & \SI{20}{km/h} & \SI{3}{km/h} & \SI{20}{km/h} \\
		\bottomrule
	\end{tabularx}
\end{table*}

\begin{table}
	\centering
	\setlength{\tabcolsep}{1ex}
	\caption{Simulation parameters.}
	\label{TABLE:Simul_Param}
	\begin{tabularx}{0.99\columnwidth}{>{\raggedright\arraybackslash}X>{\raggedright\arraybackslash}X}
		\toprule
		\textbf{Parameter} & \textbf{Value} \\
		\midrule
		Carrier frequency & \SI{28}{\GHz}\\
		System bandwidth & \SI{50}{\MHz}\\
		Subcarrier spacing & \SI{60}{\kHz}\\
		Number of subcarriers per \acs{RB} &  $12$\\
		Number of \acsp{RB} & $66$\\
		Slot duration & \SI{0.25}{\ms} \\
		OFDM symbols per slot & $14$ \\
		Channel generation procedure & As described in~\cite[Fig.7.6.4-1]{3gpp.38.901}\\
		Path loss  & Eqs. in~\cite[Table 7.4.1-1]{3gpp.38.901}\\
		Fast fading & As described in~\cite[Sec.7.5]{3gpp.38.901} and \cite[Table7.5-6]{3gpp.38.901} \\
		AWGN power per subcarrier & \SI{-174}{dBm}\\
		Noise figure &  \SI{9}{\decibel}\\
		Number of buses & $6$ \\
		Passengers~$+$~pedestrians & 72 \\
		Percentage of passengers &  $50\%$ \\
		Number of passengers per bus & $6$ \\
		\acs{CBR} packet size & $3072$ bits \\
		\acs{CBR} packet inter-arrival time & \SI{4}{slots}\\
		\bottomrule
	\end{tabularx}
\end{table}

In order to perform the simulations, two system layouts were considered. %
They were based on the Madrid grid~\cite{METIS:D6.1:2013}. 
The first layout, \FigRef{FIG:Layout-not-limited-by-interf}, was not limited by interference, while the second one, \FigRef{FIG:Layout-limited-by-interf}, was limited by interference. %

In the layout not limited by interference, \FigRef{FIG:Layout-not-limited-by-interf}, there were nine \SI{120}{m}~$\times$~\SI{120}{m} blocks. %
They were surrounded by \SI{3}{\meter}~wide sidewalks and separated of each other by \SI{14}{m}~wide streets with four lanes, two in each direction. %
In the central block there were 3 not co-located macro \acp{gNB}. %
Pedestrians and buses were initially randomly placed in the sidewalks and in the streets, respectively. %
In the intersections, they had a probability of \SI{60}{\%} to continue straight ahead\replaced{,}{ and} \SI{20}{\%} to turn left and \SI{20}{\%} to turn right. %
The pedestrians walked in the sidewalks and were allowed to cross the roads only in the intersections. %
Passengers were randomly located inside the buses at any available seat. %
During the simulation, their position relative to their bus did not change. %
In the macros-picos scenario, in addition to the three macro \acp{gNB}, there were six pico \acp{gNB} deployed as indicated in~\FigRef{FIG:Layout-not-limited-by-interf} (at the vertices of a hexagon). %
In the \ac{mIAB} scenario, \ac{mIAB} nodes were deployed at the buses. %
The \ac{DU} and \ac{MT} were placed at the back of the buses; however, the \ac{DU} was inside and the \ac{MT} was outside at the roof. %
An \ac{mIAB} node could not connect to another \ac{mIAB} node, but only be served by an \ac{IAB} donor, i.e., a \ac{gNB}. %

Concerning the layout limited by interference, i.e., \FigRef{FIG:Layout-limited-by-interf}, it had two main differences compared to the layout not limited by interference: i)~pedestrians, buses and passengers from the previous layout were concentrated in just one block, and; ii)~there was just one macro \ac{gNB} instead of three as in the previous layout. %


Regarding resource scheduling, the \ac{TDD} scheme presented in~\TabRef{TABLE:macro-pico-TDD} and standardized by \ac{3GPP} in \cite{3gpp.36.211b} was considered in the only macros and macros-picos scenarios. %
For the \ac{IAB}, in each scenario, two \ac{TDD} schemes were considered: one without \added{and other with }slots of silence\deleted{ and other with}, \added{as }presented in \TabRef{TABLE:TDD-self-interf} and \TabRef{TABLE:TDD-silent-slots}, respectively. %
The 5G-SToRM channel model~\cite{Pessoa2019} was used to model the channel of the links. %
It is an implementation of~\cite{3gpp.38.901}. %
It is spatially- and time-consistent and considers a distance-dependent path-loss, a lognormal shadowing component and small-scale fading. %
Moreover, all \deleted{the }links with the \added{\ac{IAB}}donor, i.e., \ac{IAB}donor to pedestrian, \added{\ac{mIAB}-}\ac{MT}, \added{\ac{mIAB}-}\ac{DU} and passenger, were modelled as \ac{UMa}. %
All links involving the \added{\ac{mIAB}-}\ac{MT}, except the link \added{\ac{IAB} }donor - \added{\ac{mIAB}-}\ac{MT}, were modelled as \ac{UMi}, as well as the links pedestrian to \added{\ac{mIAB}-}\ac{DU} and passenger. %
The link \ac{DU}-passenger was modelled as an indoor hotspot. %
The bus body had a penetration loss of \SI{20}{dB}~\cite{Mastrosimone2017}. %
Thus, the links between one entity inside the bus and another outside, such as \{\added{\ac{IAB} }donor - \added{\ac{mIAB}-}\ac{DU}\}, \{\added{\ac{mIAB}-}\ac{DU} - pedestrian\}, \{\added{ \ac{mIAB}-}\ac{MT} - passenger\}, and \{pedestrian - passenger\}, suffered this penetration loss. %
Furthermore, the link between a \ac{DU} of a bus and a passenger of another bus suffered twice this penetration loss. %
Besides, all \deleted{the }links that crossed the bus body were considered \ac{NLOS}. %
The other links could be either in \ac{LOS} or \ac{NLOS} with a transitional state between them as described in~\cite{3gpp.38.901}. %


\Acp{UE} and \ac{mIAB} nodes measured the \ac{RSRP} of candidate serving cells as defined in~\cite{3gpp.38.215}. %
The topology adaptation\added{, i.e., selection of \ac{IAB} donor,} was based on the highest measured \ac{RSRP}. %
Concerning the link adaptation, it was adopted the \ac{CQI}/\ac{MCS} mapping curves standardized in~\cite{3gpp.38.214} with a target \ac{BLER} of \SI{10}{\%}. %
\deleted{It was}\added{We} also considered an outer loop strategy to avoid the increase of the \ac{BLER}. %
According to this strategy, when a transmission error occurred, the estimated \ac{SINR} used for the \ac{CQI}/\ac{MCS} mapping in the link adaptation was subtracted by a back-off value of \SI{1}{dB} and, when a transmission occurred without error, the estimated \ac{SINR} had its value added by \SI{0.1}{dB}. %
Tables~\ref{TABLE:Entities-characteristics} and~\ref{TABLE:Simul_Param} present other relevant simulation parameters. 
\subsection{Simulation Results}
\label{SUBSEC:Sim_Results}

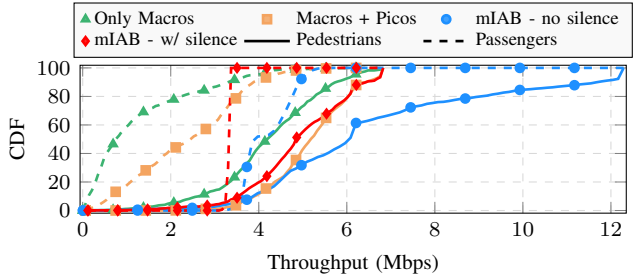
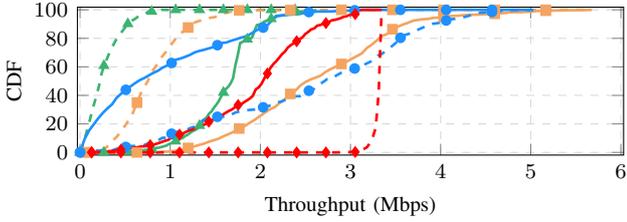
\begin{figure}
	\centering
	\subfloat[Layout not limited by interference.]{%
			\begin{tikzpicture}
		\begin{axis}[common plots axis options,
			ylabel=CDF,
			xlabel=Throughput (Mbps),
			legend style={
		    	at = {(0.5, 1.05)},
		    	anchor = south,
		    	legend columns = 3,
			}
			]
			\def\passenproportion{50}  
			\def\direction{DOWNLINK}
			\def\packetsize{2048}
			
			\def\suffix{50_DOWNLINK_bbl_20.0_spreaded_3072_bits}
			
			\pgfplotstableread [col sep=comma]
			{\plotsDataPath/e2e_thr_cdf/IAB-no_silence_\suffix.csv}\tableDataIABNoSilence
			
			\pgfplotstableread [col sep=comma]
			{\plotsDataPath/e2e_thr_cdf/IAB-with_silence_\suffix.csv}\tableDataIABWithSilence
						
			\pgfplotstableread [col sep=comma]
			{\plotsDataPath/e2e_thr_cdf/MacrosPicos_\suffix.csv} \tableDataMacrosPicos
			
			\pgfplotstableread [col sep=comma]
			{\plotsDataPath/e2e_thr_cdf/OnlyMacros_\suffix.csv} \tableDataOnlyMacros
			
			\addlegendimage{only macros style, only marks}
			\addlegendentry{Only Macros}
			\addlegendimage{macros and picos style, only marks}
			\addlegendentry{Macros + Picos}
			\addlegendimage{iab no silence style, only marks}
			\addlegendentry{\acs{mIAB} - no silence}
			\addlegendimage{iab with silence style, only marks}
			\addlegendentry{\acs{mIAB} - w/ silence}
			
			\addlegendimage{pedestrian style, black}
			\addlegendentry{Pedestrians}
			\addlegendimage{passenger style, black}
			\addlegendentry{Passengers}
			
			\addplot[only macros style, pedestrian style]
			table[x=x, y=ped_y] from \tableDataOnlyMacros;
			
			\addplot[only macros style, passenger style]
			table[x=x, y=passen_y] from \tableDataOnlyMacros;
			
			\addplot[macros and picos style, pedestrian style]
			table[x=x, y=ped_y] from \tableDataMacrosPicos;
			
			\addplot[macros and picos style, passenger style]
			table[x=x, y=passen_y] from \tableDataMacrosPicos;
						
			\addplot[iab no silence style, pedestrian style]
			table[x=x, y=ped_y] from \tableDataIABNoSilence;
			
			\addplot[iab no silence style, passenger style]
			table[x=x, y=passen_y] from \tableDataIABNoSilence;
			
			\addplot[iab with silence style, pedestrian style]
			table[x=x, y=ped_y] from \tableDataIABWithSilence;
			
			\addplot[iab with silence style, passenger style]
			table[x=x, y=passen_y] from \tableDataIABWithSilence;
		\end{axis}
	\end{tikzpicture}
	
		\label{FIG:DL-Throughput-not-limited-by-interference}
	}

	\subfloat[Layout limited by interference.]{%
			\begin{tikzpicture}
		\begin{axis}[common plots axis options,
			ylabel=CDF,
			xlabel=Throughput (Mbps),
			xmax = 6,
			legend style={
		    	at = {(0.5, 1.05)},
		    	anchor = south,
		    	legend columns = 3,
			}
			]
			\def\passenproportion{50}  
			\def\direction{DOWNLINK}
			\def\packetsize{2048}
			
			\def\suffix{50_DOWNLINK_bbl_20.0_hotspot_3072_bits}
			
			\pgfplotstableread [col sep=comma]
			{\plotsDataPath/e2e_thr_cdf/IAB-no_silence_\suffix.csv}\tableDataIABNoSilence
			
			\pgfplotstableread [col sep=comma]
			{\plotsDataPath/e2e_thr_cdf/IAB-with_silence_\suffix.csv}\tableDataIABWithSilence
						
			\pgfplotstableread [col sep=comma]
			{\plotsDataPath/e2e_thr_cdf/MacrosPicos_\suffix.csv} \tableDataMacrosPicos
			
			\pgfplotstableread [col sep=comma]
			{\plotsDataPath/e2e_thr_cdf/OnlyMacros_\suffix.csv} \tableDataOnlyMacros
			
%
			
			\addplot[only macros style, pedestrian style]
			table[x=x, y=ped_y] from \tableDataOnlyMacros;
			
			\addplot[only macros style, passenger style]
			table[x=x, y=passen_y] from \tableDataOnlyMacros;
			
			\addplot[macros and picos style, pedestrian style]
			table[x=x, y=ped_y] from \tableDataMacrosPicos;
			
			\addplot[macros and picos style, passenger style]
			table[x=x, y=passen_y] from \tableDataMacrosPicos;
						
			\addplot[iab no silence style, pedestrian style]
			table[x=x, y=ped_y] from \tableDataIABNoSilence;
			
			\addplot[iab no silence style, passenger style]
			table[x=x, y=passen_y] from \tableDataIABNoSilence;
			
			\addplot[iab with silence style, pedestrian style]
			table[x=x, y=ped_y] from \tableDataIABWithSilence;
			
			\addplot[iab with silence style, passenger style]
			table[x=x, y=passen_y] from \tableDataIABWithSilence;
		\end{axis}
	\end{tikzpicture}
	
		\label{FIG:DL-Throughput-limited-by-interference}
	}
	\caption{\added{The \ac{CDF} of }\ac{DL} throughput\added{ for different interferce scenarios}.}
	\label{FIG:DL-Throughput}	
\end{figure}

\deleted{Firstly, let}\added{Let} us analyze the impact of introducing \ac{mIAB} with and without slots of silence on the \ac{DL} throughput of passengers and pedestrians presented in \FigRef{FIG:DL-Throughput}. %
More specifically, \FigRef{FIG:DL-Throughput-not-limited-by-interference} and \FigRef{FIG:DL-Throughput-limited-by-interference} refer to the scenarios not limited and limited by interference, respectively. %


First, notice in \FigRef{FIG:DL-Throughput-not-limited-by-interference} that, in the scenario not limited by interference, deploying \ac{mIAB} nodes with and without slots of silence outstandingly improved the passengers' \ac{DL} throughput compared to the cases of only macros and macros-picos. %
While in the only macros and macros-picos only $11\%$ and $31\%$, respectively, of the passengers had \ac{DL} throughput higher than \SI{3.2}{MBits/s}, with the \ac{mIAB} almost all the passengers had \ac{DL} throughput higher than \SI{3.2}{MBits/s}. %
Moreover, since in this scenario the interference was not a problem, having slots of silence was a waste of resources. %
Hence, the \ac{mIAB} without slots of silence, i.e., always available for transmission, outperformed the case of \ac{mIAB} with slots of silence. %

In the scenario limited by interference, \FigRef{FIG:DL-Throughput-limited-by-interference}, passengers were also benefited by the deployment of \ac{mIAB} nodes with or without slots of silence. %
However, in this scenario, \ac{mIAB} without slots of silence harmed more the access links of pedestrians than those of passengers. %
This occurred because the passengers were close to the serving antennas, \deleted{which reduced the negative}\added{reducing the} impact of interference, while the pedestrians were further to their serving \ac{gNB} than to the interference source, i.e., passengers transmitting in the \ac{UL} according to \TabRef{TABLE:TDD-silent-slots}. %

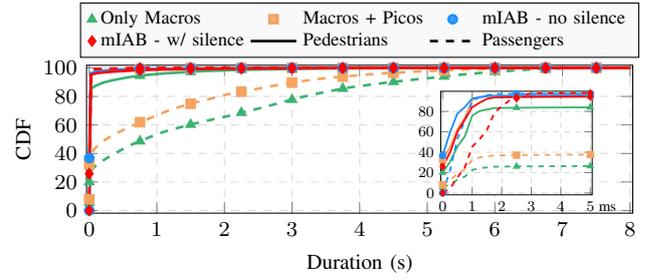
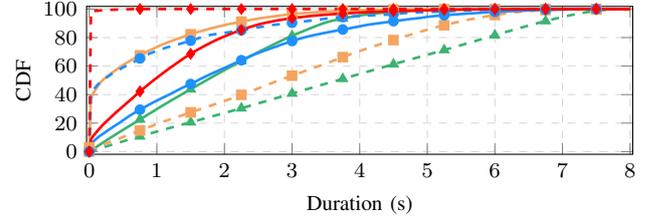
\begin{figure}
	\centering
	\subfloat[Layout not limited by interference.]{%
			\begin{tikzpicture}
		\begin{axis}[common plots axis options,
			xlabel={Duration (s)},
			ylabel=CDF,
			xmax = 8,
			xtick = {0,1,...,8},
			legend style={
				at = {(0.5, 1.05)},
				anchor = south,
				legend columns = 3,
			}
			]
			
			\def\suffix{50_DOWNLINK_bbl_20.0_spreaded_3072_bits}
			
			\pgfplotstableread [col sep=comma]
			{\plotsDataPath/packet_latency_cdf/IAB-no_silence_\suffix.csv}\tableDataIABNoSilence
			
			\pgfplotstableread [col sep=comma]
			{\plotsDataPath/packet_latency_cdf/IAB-with_silence_\suffix.csv}\tableDataIABWithSilence
			
			\pgfplotstableread [col sep=comma]
			{\plotsDataPath/packet_latency_cdf/MacrosPicos_\suffix.csv} \tableDataMacrosPicos
			
			\pgfplotstableread [col sep=comma]
			{\plotsDataPath/packet_latency_cdf/OnlyMacros_\suffix.csv} \tableDataOnlyMacros
						
			\addlegendimage{only macros style, only marks}
			\addlegendentry{Only Macros}
			\addlegendimage{macros and picos style, only marks}
			\addlegendentry{Macros + Picos}
			\addlegendimage{iab no silence style, only marks}
			\addlegendentry{\acs{mIAB} - no silence}
			\addlegendimage{iab with silence style, only marks}
			\addlegendentry{\acs{mIAB} - w/ silence}
			
			\addlegendimage{pedestrian style, black}
			\addlegendentry{Pedestrians}
			\addlegendimage{passenger style, black}
			\addlegendentry{Passengers}
			
			\def\markrepeatstep{30}
			
			\addplot[only macros style, pedestrian style,  mark repeat={\markrepeatstep}]
			table[x=latency_ms_cdf_x, y=pedestrians_latency_cdf_y] from \tableDataOnlyMacros;
			
			\addplot[only macros style, passenger style,  mark repeat={\markrepeatstep}]
			table[x=latency_ms_cdf_x, y=passengers_latency_cdf_y] from \tableDataOnlyMacros;
			
			\addplot[macros and picos style, pedestrian style,  mark repeat={\markrepeatstep}]
			table[x=latency_ms_cdf_x, y=pedestrians_latency_cdf_y] from \tableDataMacrosPicos;
			
			\addplot[macros and picos style, passenger style,  mark repeat={\markrepeatstep}]
			table[x=latency_ms_cdf_x, y=passengers_latency_cdf_y] from \tableDataMacrosPicos;
			
			\addplot[iab no silence style, pedestrian style,  mark repeat={\markrepeatstep}]
			table[x=latency_ms_cdf_x, y=pedestrians_latency_cdf_y] from \tableDataIABNoSilence;
			
			\addplot[iab no silence style, passenger style,  mark repeat={\markrepeatstep}]
			table[x=latency_ms_cdf_x, y=passengers_latency_cdf_y] from \tableDataIABNoSilence;
			
			\addplot[iab with silence style, pedestrian style,  mark repeat={\markrepeatstep}]
			table[x=latency_ms_cdf_x, y=pedestrians_latency_cdf_y] from \tableDataIABWithSilence;
			
			\addplot[iab with silence style, passenger style,  mark repeat={\markrepeatstep}]
			table[x=latency_ms_cdf_x, y=passengers_latency_cdf_y] from \tableDataIABWithSilence;
			
			\coordinate (pt) at (axis cs:5.2,11);
		\end{axis}
	
		\begin{scope}[scale=0.7, transform shape]
			\begin{axis}[common plots axis options,
				at={(pt)},
				axis background/.style={fill=white},
				width=0.5\columnwidth,
				height=0.4\columnwidth,
				xlabel={ms},
				every axis x label/.style={
					at={(ticklabel* cs:1)},
					anchor=north west,
					font=\footnotesize,
					yshift=-1.5pt
				},
				xmax = 5,
				xtick = {0,1,...,5},
				]
				
				\def\suffix{50_DOWNLINK_bbl_20.0_spreaded_3072_bits}
				
				\pgfplotstableread [col sep=comma]
				{\plotsDataPath/packet_latency_cdf/zoom/IAB-no_silence_\suffix.csv}\tableDataIABNoSilence
				
				\pgfplotstableread [col sep=comma]
				{\plotsDataPath/packet_latency_cdf/zoom/IAB-with_silence_\suffix.csv}\tableDataIABWithSilence
				
				\pgfplotstableread [col sep=comma]
				{\plotsDataPath/packet_latency_cdf/zoom/MacrosPicos_\suffix.csv} \tableDataMacrosPicos
				
				\pgfplotstableread [col sep=comma]
				{\plotsDataPath/packet_latency_cdf/zoom/OnlyMacros_\suffix.csv} \tableDataOnlyMacros
				
				\addplot[only macros style, pedestrian style]
				table[x expr=\thisrow{latency_ms_cdf_x}*1000, y=pedestrians_latency_cdf_y] from \tableDataOnlyMacros;
				
				\addplot[only macros style, passenger style]
				table[x expr=\thisrow{latency_ms_cdf_x}*1000, y=passengers_latency_cdf_y] from \tableDataOnlyMacros;
				
				\addplot[macros and picos style, pedestrian style]
				table[x expr=\thisrow{latency_ms_cdf_x}*1000, y=pedestrians_latency_cdf_y] from \tableDataMacrosPicos;
				
				\addplot[macros and picos style, passenger style]
				table[x expr=\thisrow{latency_ms_cdf_x}*1000, y=passengers_latency_cdf_y] from \tableDataMacrosPicos;
				
				\addplot[iab no silence style, pedestrian style]
				table[x expr=\thisrow{latency_ms_cdf_x}*1000, y=pedestrians_latency_cdf_y] from \tableDataIABNoSilence;
				
				\addplot[iab no silence style, passenger style]
				table[x expr=\thisrow{latency_ms_cdf_x}*1000, y=passengers_latency_cdf_y] from \tableDataIABNoSilence;
				
				\addplot[iab with silence style, pedestrian style]
				table[x expr=\thisrow{latency_ms_cdf_x}*1000, y=pedestrians_latency_cdf_y] from \tableDataIABWithSilence;
				
				\addplot[iab with silence style, passenger style]
				table[x expr=\thisrow{latency_ms_cdf_x}*1000, y=passengers_latency_cdf_y] from \tableDataIABWithSilence;
			\end{axis}
		\end{scope}
	\end{tikzpicture}
	
		\label{FIG:DL-Latency-not-limited-by-interference}
	}

	\subfloat[Layout limited by interference.]{%
			\begin{tikzpicture}
		\begin{axis}[common plots axis options,
			xlabel={Duration (s)},
			ylabel=CDF,
			xmax = 8,
			xtick = {0,1,...,8},
			legend style={
				at = {(0.5, 1.05)},
				anchor = south,
				legend columns = 3,
			}
			]
			
			\def\suffix{50_DOWNLINK_bbl_20.0_hotspot_3072_bits}
			
			\pgfplotstableread [col sep=comma]
			{\plotsDataPath/packet_latency_cdf/IAB-no_silence_\suffix.csv}\tableDataIABNoSilence
			
			\pgfplotstableread [col sep=comma]
			{\plotsDataPath/packet_latency_cdf/IAB-with_silence_\suffix.csv}\tableDataIABWithSilence
			
			\pgfplotstableread [col sep=comma]
			{\plotsDataPath/packet_latency_cdf/MacrosPicos_\suffix.csv} \tableDataMacrosPicos
			
			\pgfplotstableread [col sep=comma]
			{\plotsDataPath/packet_latency_cdf/OnlyMacros_\suffix.csv} \tableDataOnlyMacros
						
%
			
			\def\markrepeatstep{30}
			
			\addplot[only macros style, pedestrian style,  mark repeat={\markrepeatstep}]
			table[x=latency_ms_cdf_x, y=pedestrians_latency_cdf_y] from \tableDataOnlyMacros;
			
			\addplot[only macros style, passenger style,  mark repeat={\markrepeatstep}]
			table[x=latency_ms_cdf_x, y=passengers_latency_cdf_y] from \tableDataOnlyMacros;
			
			\addplot[macros and picos style, pedestrian style,  mark repeat={\markrepeatstep}]
			table[x=latency_ms_cdf_x, y=pedestrians_latency_cdf_y] from \tableDataMacrosPicos;
			
			\addplot[macros and picos style, passenger style,  mark repeat={\markrepeatstep}]
			table[x=latency_ms_cdf_x, y=passengers_latency_cdf_y] from \tableDataMacrosPicos;
			
			\addplot[iab no silence style, pedestrian style,  mark repeat={\markrepeatstep}]
			table[x=latency_ms_cdf_x, y=pedestrians_latency_cdf_y] from \tableDataIABNoSilence;
			
			\addplot[iab no silence style, passenger style,  mark repeat={\markrepeatstep}]
			table[x=latency_ms_cdf_x, y=passengers_latency_cdf_y] from \tableDataIABNoSilence;
			
			\addplot[iab with silence style, pedestrian style,  mark repeat={\markrepeatstep}]
			table[x=latency_ms_cdf_x, y=pedestrians_latency_cdf_y] from \tableDataIABWithSilence;
			
			\addplot[iab with silence style, passenger style,  mark repeat={\markrepeatstep}]
			table[x=latency_ms_cdf_x, y=passengers_latency_cdf_y] from \tableDataIABWithSilence;
			
		\end{axis}
	\end{tikzpicture}
	
		\label{FIG:DL-Latency-limited-by-interference}
	}
	\caption{\Ac{DL} latency.}
	\label{FIG:DL-Latency}	
\end{figure}

Regarding the \ac{DL} latency, \FigRef{FIG:DL-Latency} presents the \ac{CDF} of \ac{DL} latency for pedestrians and passengers in all considered cases. %
Notice that, in the scenario not limited by interference, \FigRef{FIG:DL-Latency-not-limited-by-interference}, on the one hand \ac{mIAB} presented \ac{DL} latency lower than \SI{5}{ms} for at least $90\%$ of both pedestrians and passengers. %
On the other hand, the only macros and macros-picos cases presented \ac{DL} latency greater than \SI{0.5}{s} for the majority of the passengers, with some of them facing a delay higher than \SI{5}{s}. %
This fact highlights that having only macros and/or picos is not enough to provide a good connection to onboard passengers. %


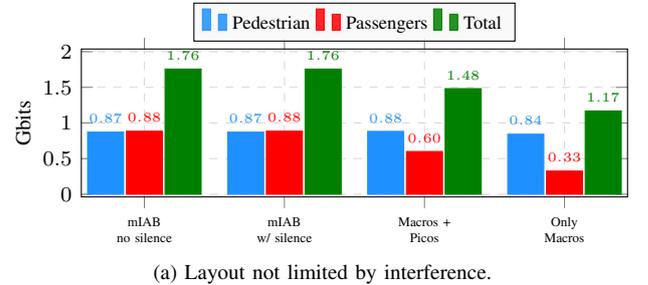
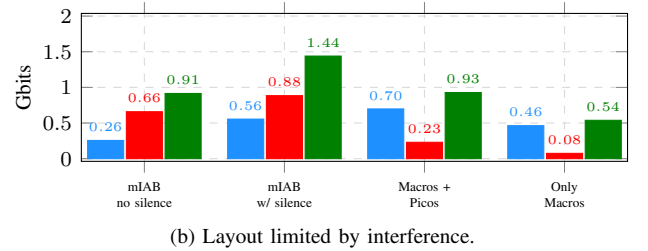
\begin{figure}
	\centering
	\subfloat[Layout not limited by interference.]{%
			\begin{tikzpicture}
		\begin{axis}[bar axis options,,
			ylabel={Gbits},
			ymin=0, ymax=2,
			xtick=data,
			xticklabels={\acs{mIAB}\\no silence, \acs{mIAB}\\w/ silence, Macros +\\Picos, Only\\Macros},
			x tick label style={
				font=\tiny,
				align=center,
			},
			legend style={
				at = {(0.5, 1.05)},
				anchor = south,
				legend columns = 3,
			},
			bar width=3ex,
			enlarge x limits=0.15,
			nodes near coords,
			every node near coord/.append style={
				font=\tiny,
				/pgf/number format/fixed,
				/pgf/number format/fixed zerofill,
				/pgf/number format/precision=2
			}
			]			
			\pgfplotstableread [col sep=comma]
			{\plotsDataPath/total_tx_bits/50_DOWNLINK_bbl_20.0_spreaded_3072_bits.csv}\tableData
			
			\addlegendimage{pedestrian bar style}
			\addlegendentry{Pedestrian}
			\addlegendimage{passenger bar style}
			\addlegendentry{Passengers}
			\addlegendimage{aggregated bar style}
			\addlegendentry{Total}
			
			\addplot[pedestrian bar style]
			table[x=idx, y expr = \thisrow{num_mbits_ped}*1e-3] from \tableData;
			
			\addplot[passenger bar style]
			table[x=idx, y expr = \thisrow{num_mbits_passen}*1e-3] from \tableData;
			
			\addplot[aggregated bar style]
			table[x=idx, y expr = \thisrow{num_mbits_aggreg}*1e-3] from \tableData;
			
		\end{axis}
	\end{tikzpicture}
	
		\label{FIG:Tot-transmitted-bits-not-limited-by-interference}
	}

	\subfloat[Layout limited by interference.]{%
			\begin{tikzpicture}
		\begin{axis}[bar axis options,,
			ylabel={Gbits},
			ymin=0, ymax=2,
			xtick=data,
			xticklabels={\acs{mIAB}\\no silence, \acs{mIAB}\\w/ silence, Macros +\\Picos, Only\\Macros},
			x tick label style={
				font=\tiny,
				align=center,
			},
			legend style={
				at = {(0.5, 1.05)},
				anchor = south,
				legend columns = 3,
			},
			bar width=3ex,
			enlarge x limits=0.15,
			nodes near coords,
			every node near coord/.append style={
				font=\tiny,
				/pgf/number format/fixed,
				/pgf/number format/fixed zerofill,
				/pgf/number format/precision=2
			}
			]			
			\pgfplotstableread [col sep=comma]
			{\plotsDataPath/total_tx_bits/50_DOWNLINK_bbl_20.0_hotspot_3072_bits.csv}\tableData
			
			
			\addplot[pedestrian bar style]
			table[x=idx, y expr = \thisrow{num_mbits_ped}*1e-3] from \tableData;
			
			\addplot[passenger bar style]
			table[x=idx, y expr = \thisrow{num_mbits_passen}*1e-3] from \tableData;
			
			\addplot[aggregated bar style]
			table[x=idx, y expr = \thisrow{num_mbits_aggreg}*1e-3] from \tableData;
			
		\end{axis}
	\end{tikzpicture}
	
		\label{FIG:Tot-transmitted-bits-limited-by-interference}
	}
	\caption{Total transmitted bits.}
	\label{FIG:Tot-transmitted-bits}	
\end{figure}

Figure~\ref{FIG:Tot-transmitted-bits} presents the average of the total amount of bits transmitted in the \ac{DL} by each solution. %
Since there were $36$~passengers and $36$~pedestrians, the amount of generated bits in the \ac{DL} to passengers and pedestrians was equal to:

\footnotesize 
\begin{multline} \label{EQ:total_bits}
36\text{ UEs}\left(\frac{1\text{ packet}}{4\text{ slots}}\right )\left(\frac{1\text{ slot}}{0.25\text{ ms}}\right)\left(\frac{8,000\text{ ms}}{\text{simulation}}\right)\left(\frac{3,072\text{ bits}}{1\text{ UE}\times 1\text{ packet}}\right) \\
=884.7\text{ Mbits/simulation}.
\end{multline}
\normalsize

First, notice that, even though in the scenario not limited by interference the \ac{mIAB} with slots of silence presented lower \ac{DL} throughput compared to the \ac{mIAB} without slots of silence, as shown in~\FigRef{FIG:DL-Throughput-not-limited-by-interference}, both solutions were able to deliver almost all the generated data to pedestrians and passengers, as presented in~\FigRef{FIG:Tot-transmitted-bits-not-limited-by-interference}. %
Moreover, \deleted{when }comparing how their performances changed when switching from a scenario not limited by interference, \FigRef{FIG:Tot-transmitted-bits-not-limited-by-interference}, to a scenario limited by interference, \FigRef{FIG:Tot-transmitted-bits-limited-by-interference}, we can see that: i)~the \ac{mIAB} without slots of silence was severely impacted by the interference, and; ii)~the pedestrians were more affected than the passengers. %
\deleted{As already mentioned, this}\added{This} occurred because the passengers were close to the transmitting antennas, \deleted{which reduced the negative}\added{reducing the} impact of interference, while the pedestrians were closer to the interference source than to the serving \ac{gNB}. %
Important to remark that in the layout limited by interference, the passengers received less than $10\%$ and $26\%$ of the generated data in the only macros and macro-picos scenarios, respectively. %

\begin{figure}
	\centering	
	\subfloat[Layout not limited by interference and \ac{TDD} scheme without slot of silence.]{%
			\begin{tikzpicture}
		\begin{axis}[mcs axis options,
			ylabel=CDF,
			ymax = 30,
			ytick = {0,10,...,30},
			legend style={
				legend pos=north west,
			}
			]
			\def\passenproportion{50}  
			\def\packetsize{3072}
			\def\direction{DOWNLINK}
			\def\linktype{Backhaul}
			\def\scenario{IAB-no_silence}
			
			\pgfplotstableread [col sep=comma]
			{\plotsDataPath/mcs_hist/\scenario_\passenproportion_bbl_20.0_spreaded_\packetsize_bits.csv}\tableData
			
			\addlegendimage{ack bar style}
			\addlegendentry{ACKs}
			\addlegendimage{nack bar style}
			\addlegendentry{NACKs}
			
			\draw[black, opacity=0.4, fill=blue,sharp plot,dashed] (-1, 10) -- (16, 10) ;
			\fill[gray, opacity=0.1] (-1, -1) -- (-1, 10) -- (16, 10) -- (16, -1) -- cycle;
			
			\addplot[ack bar style]
			table[x=MCS, y=perc_\direction_\linktype_acks] from \tableData;
			
			\addplot[nack bar style]
			table[x=MCS, y=perc_\direction_\linktype_nacks] from \tableData;
			
		\end{axis}
	\end{tikzpicture}
	
		\label{FIG:backhaul-MCS-usage-not-limited-by-interf-without-silence}
	}

	\subfloat[Layout not limited by interference and \ac{TDD} scheme with slot of silence.]{%
			\begin{tikzpicture}
		\begin{axis}[mcs axis options,
			ylabel=CDF,
			ymax = 40,
			ytick = {0,10,...,40},
			legend style={
				legend pos=north west,
			}
			]
			\def\passenproportion{50}  
			\def\packetsize{3072}
			\def\direction{DOWNLINK}
			\def\linktype{Backhaul}
			\def\scenario{IAB-with_silence}
			
			\pgfplotstableread [col sep=comma]
			{\plotsDataPath/mcs_hist/\scenario_\passenproportion_bbl_20.0_spreaded_\packetsize_bits.csv}\tableData
			
			\addlegendimage{ack bar style}
			\addlegendentry{ACKs}
			\addlegendimage{nack bar style}
			\addlegendentry{NACKs}
			
			\draw[black, opacity=0.4, fill=blue,sharp plot,dashed] (-1, 10) -- (16, 10) ;
			\fill[gray, opacity=0.1] (-1, -1) -- (-1, 10) -- (16, 10) -- (16, -1) -- cycle;
			
			\addplot[ack bar style]
			table[x=MCS, y=perc_\direction_\linktype_acks] from \tableData;
			
			\addplot[nack bar style]
			table[x=MCS, y=perc_\direction_\linktype_nacks] from \tableData;
			
		\end{axis}
	\end{tikzpicture}
	
		\label{FIG:backhaul-MCS-usage-not-limited-by-interf-with-silence}
	}

	\subfloat[Layout limited by interference and \ac{TDD} scheme without slot of silence.]{%
			\begin{tikzpicture}
		\begin{axis}[mcs axis options,
			ylabel=CDF,
			ymax = 15,
			ytick = {0,5,...,15},
			legend style={
				legend pos=north east,
			}
			]
			\def\passenproportion{50}  
			\def\packetsize{3072}
			\def\direction{DOWNLINK}
			\def\linktype{Backhaul}
			\def\scenario{IAB-no_silence}
			
			\pgfplotstableread [col sep=comma]
			{\plotsDataPath/mcs_hist/\scenario_\passenproportion_bbl_20.0_hotspot_\packetsize_bits.csv}\tableData
			
			\addlegendimage{ack bar style}
			\addlegendentry{ACKs}
			\addlegendimage{nack bar style}
			\addlegendentry{NACKs}
			
			\draw[black, opacity=0.4, fill=blue,sharp plot,dashed] (-1, 10) -- (16, 10) ;
			\fill[gray, opacity=0.1] (-1, -1) -- (-1, 10) -- (16, 10) -- (16, -1) -- cycle;
			
			\addplot[ack bar style]
			table[x=MCS, y=perc_\direction_\linktype_acks] from \tableData;
			
			\addplot[nack bar style]
			table[x=MCS, y=perc_\direction_\linktype_nacks] from \tableData;
			
		\end{axis}
	\end{tikzpicture}
	
		\label{FIG:backhaul-MCS-usage-limited-by-interf-without-silence}
	}

	\subfloat[Layout limited by interference and \ac{TDD} scheme with slot of silence.]{%
			\begin{tikzpicture}
		\begin{axis}[mcs axis options,
			ylabel=CDF,
			ymax = 70,
			ytick = {0,10,...,70},
			legend style={
				legend pos=north west,
			}
			]
			\def\passenproportion{50}  
			\def\packetsize{3072}
			\def\direction{DOWNLINK}
			\def\linktype{Backhaul}
			\def\scenario{IAB-with_silence}
			
			\pgfplotstableread [col sep=comma]
			{\plotsDataPath/mcs_hist/\scenario_\passenproportion_bbl_20.0_hotspot_\packetsize_bits.csv}\tableData
			
			\addlegendimage{ack bar style}
			\addlegendentry{ACKs}
			\addlegendimage{nack bar style}
			\addlegendentry{NACKs}
			
			\draw[black, opacity=0.4, fill=blue,sharp plot,dashed] (-1, 10) -- (16, 10) ;
			\fill[gray, opacity=0.1] (-1, -1) -- (-1, 10) -- (16, 10) -- (16, -1) -- cycle;
			
			\addplot[ack bar style]
			table[x=MCS, y=perc_\direction_\linktype_acks] from \tableData;
			
			\addplot[nack bar style]
			table[x=MCS, y=perc_\direction_\linktype_nacks] from \tableData;
			
		\end{axis}
	\end{tikzpicture}
	
		\label{FIG:backhaul-MCS-usage-limited-by-interf-with-silence}
	}
	
	\caption{Backhaul \ac{MCS} usage.}
	\label{FIG:backhaul-MCS-usage}	
\end{figure}
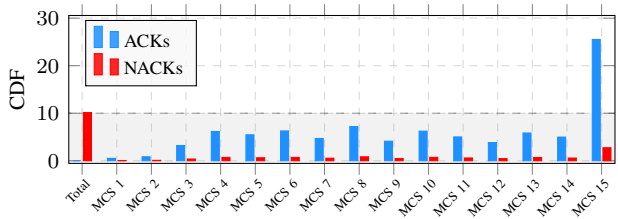
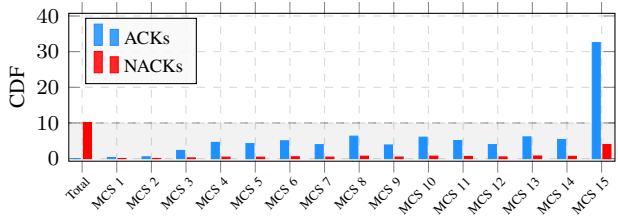
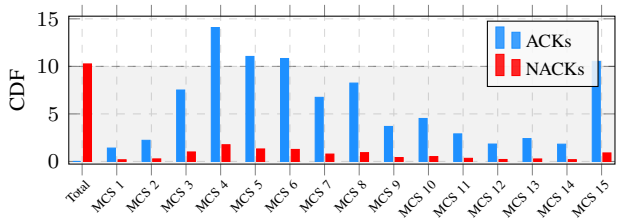
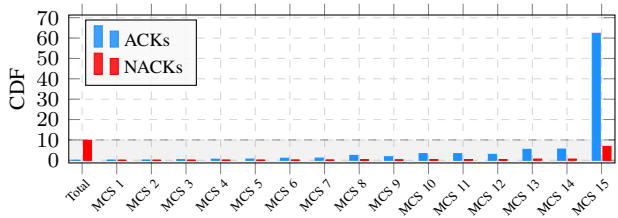

Until now, the analyses focused on the access links. %
Let us now analyze\added{, in \FigRef{FIG:backhaul-MCS-usage},} \deleted{the backhaul link of the \ac{mIAB} nodes. %
Figure~\ref{FIG:backhaul-MCS-usage} presents }the histogram of \ac{MCS} usage\added{ in the backhaul link of the \ac{mIAB} nodes}. %
Figures~\ref{FIG:backhaul-MCS-usage-not-limited-by-interf-without-silence} and ~\ref{FIG:backhaul-MCS-usage-not-limited-by-interf-with-silence} concern the cases without and with slots of silence, respectively, in the scenario not limited by interference. %
Comparing these two figures, we can see that in this scenario both solutions were able to transmit data in the backhaul using the highest \ac{MCS}. %
As expected, in the layout limited by interference, \ac{mIAB} without slots of silence, \FigRef{FIG:backhaul-MCS-usage-limited-by-interf-without-silence}, had its backhaul affected by the interference impacting the used \acp{MCS}. %

\deleted{Interesting to compare}\added{Comparing} Figs.~\ref{FIG:backhaul-MCS-usage-not-limited-by-interf-with-silence} and~\ref{FIG:backhaul-MCS-usage-limited-by-interf-with-silence}\deleted{. %
Notice}\added{, notice} that for the \ac{mIAB} solution with slots of silence, when we moved from the layout not limited by interference to the scenario limited by interference, the quality of the backhaul improved. %
This is due to the fact that when the backhaul was in \ac{DL}, i.e., the \added{\ac{mIAB}-}\ac{MT} \deleted{part of the \ac{mIAB} node }was receiving, the main interference came from the onboard \acp{UE} transmitting in the \ac{UL} to the \added{\ac{mIAB}-}\ac{DU}\deleted{ part of the \ac{mIAB} node}. %
The distance between the \added{\ac{mIAB}-}\ac{MT} and the onboard \acp{UE} (source of interference) was approximately constant, thus changing the scenario did not considerably change the level of interference in the backhaul. %
However, in the scenario limited by interference, the \ac{mIAB} nodes were closer to the serving \ac{IAB} donor, which increased the strength of the received signal. %


\section{Conclusions}
\label{SEC:Conclusion}

We concluded that \ac{mIAB} \deleted{outstandingly improves}\added{has the potential to improve} the \deleted{connection}\added{throughput and latency} of \deleted{onboard }\acp{UE}\added{ onboard of busses}, which poorly \deleted{perform}\added{performed} in \added{the considered benchmark }scenarios\deleted{ without \ac{mIAB}}, i.e., with only macro and/or pico \acp{gNB}. %
More specifically, \ac{mIAB} with a \ac{TDD} scheme with slots of silence \deleted{is a good option for}\added{outperformed benchmark solutions in} scenarios \added{with different levels of interference}\deleted{limited or not by interference}. %
Furthermore, in \added{a considered}\deleted{a} scenario\added{, which was} not limited by interference, a \ac{TDD} scheme without slots of silence \deleted{performs}\added{performed} even better. %
Thus, since a \ac{TDD} scheme with slots of silence performed well in both scenarios, it could be adopted as a default solution. %
\added{Moreover, depending on the data traffic, geographical conditions, etc. it may be useful to adapt the \ac{TDD} scheme}. %
\deleted{As perspectives for this work, based on well defined triggering conditions (e.g., \ac{mIAB} node enters a geographic area known as not limited by interference) the \ac{TDD} scheme could be switched to a scheme without the slots of silence to improve the system performance.} %

\printbibliography
\end{document}